\documentclass
  [letterpaper,11pt,abstracton,normalheadings,tbtags]{scrartcl} 
\usepackage[margin=1in]{geometry}

\usepackage{graphicx}

\usepackage[cmex10]{empheq}
\usepackage{braket,amssymb}
\usepackage{dsfont,mathrsfs}
\usepackage[standard,amsmath,thmmarks] {ntheorem}

\newtheorem {problem}       {Problem}        
\newtheorem*  {structure}         {Structural Result}
\newtheorem*  {first-structure}   {First Structural Result}
\newtheorem*  {second-structure}  {Second Structural Result}

\let\ALPHABET \mathcal

\let\DEFINED        \colonequals
\let\BYDEFINITION   \equalscolon

\usepackage{mathsets}
\newcommand     \IND   {\mathds{1}}
\definemathset  [EXP]  {\mathds{E}}  \{ \vert \}
\definemathset  [PR]   {\mathds{P}}  ( \vert )
\definemathset  [PSP]  {\mathcal{P}}  \{ \vert \}
\newcommand\PPR{\mathds{P}}

\newcommand\reals{\mathds{R}}

\title
   {Optimal Control Strategies in Delayed Sharing Information Structures}

\author 
   {Ashutosh Nayyar, Aditya Mahajan and Demosthenis Teneketzis}

\date
  {February 12, 2010}

\begin{document}

\maketitle

\begin{abstract}
  The $n$-step delayed sharing information structure is investigated. This
  information structure comprises of $K$ controllers  that share their
  information with a delay of $n$ time steps. This information structure is a
  link between the classical information structure, where information is shared
  perfectly between the controllers, and a non-classical information structure,
  where there is no ``lateral'' sharing of information among the controllers.
  Structural results for optimal control strategies for systems with such
  information structures are presented. A sequential methodology for finding the
  optimal strategies is also derived.  The solution approach provides an insight
  for identifying structural results and sequential decomposition for general
  decentralized stochastic control problems.
\end{abstract}

\section{Introduction}

\subsection {Motivation}

One of the difficulties in optimal design of decentralized control systems is
handling the increase of data at the control stations with time. This increase
in data means that the domain of control laws increases with time which, in
turn, creates two difficulties. Firstly, the number of control strategies
increases doubly exponentially with time; this makes it harder to search for an
optimal strategy. Secondly, even if an optimal strategy is found, implementing
functions with time increasing domain is difficult. 

In centralized stochastic control~\cite{KumarVaraiya:1986}, these difficulties
can be circumvented by using the conditional probability of the state given the
data available at the control station as a sufficient statistic (where the data
available to a control station comprises of all observations and control actions
till the current time) . This conditional probability, called \emph{information
state}, takes values in a time-invariant space. Consequently, we can restrict
attention to control laws with time-invariant domain. Such results, in which data
that is increasing with time is ``compressed'' to a sufficient statistic taking
values in a time-invariant space, are called \emph{structural results}. While
the information state and structural result for centralized stochastic control
problems are well known, no general methodology to find such information states
or structural results exists for decentralized stochastic control problems.

The structural results in centralized stochastic control are related to the
concept of separation. In centralized stochastic control, the information state,
which is conditional probability of the state given all the available data, does
not depend on the control strategy (which is the collection of control laws used
at different time instants). This has been called a one-way separation between
estimation and control. An important consequence of this separation is that for
any given choice of control laws till time $t-1$ and a given realization of the
system variables till time $t$, the information states at future times do not
depend on the choice of the control law at time $t$ but only on the realization
of control action at time $t$. Thus, the future information states are
\emph{separated} from the choice of the current control law. This  fact is
crucial for the formulation of the classical dynamic program where at each step
the optimization problem is to find the best control action for a given
realization of the information state. No analogous separation results are known
for general decentralized systems. 

In this paper, we find structural results for decentralized control systems with
delayed sharing information structures. In a system with $n$-step delayed
sharing, every control station knows the $n$-step prior observations and control
actions of all other control stations. This information structure, proposed by
Witsenhausen in~\cite{Witsenhausen:1971}, is a link between the classical
information structures, where information is shared perfectly among the
controllers, and the non-classical information structures, where there is no
``lateral'' sharing of information among the controllers. In his seminal
paper~\cite{Witsenhausen:1971},  Witsenhausen asserted a structural result for
this model without any proof. Varaiya and Walrand~\cite{WalrandVaraiya:1978}
proved that Witsenhausen's assertion was true for $n=1$ but false for $n>1$. For
$n>1$, Kurtaran~\cite{Kurtaran:1979} proposed another structural result.
However, Kurtaran proved his result only for the terminal time step (that is,
the last time step in a finite horizon problem); for non-terminal time steps, he
gave an abbreviated argument, which we believe is incomplete. (The details are
given in Section~\ref{sec:kurtaran} of the paper).

We prove two structural results of the optimal control laws for the delayed
sharing information structure. We compare our results to those conjectured by
Witsenhausen and show that our structural results for $n$-step delay sharing
information structure simplify to that of Witsenhausen for $n=1$; for $n>1$, our
results are different from the result proposed by Kurtaran.

Our structural results do not have the separated nature of
centralized stochastic control: for any given realization of the system
variables till time $t$, the realization of information states at future times
depend on the choice of the control law at time $t$. 
However, our second structural result shows that this dependence only propagates
to the next $n-1$ time steps. Thus, the information states from time $t+n-1$
onwards are separated from the choice of control laws before time $t$; they only
depend on the realization of control actions at time $t$. We call this a
\emph{delayed} separation between information states and control laws. 

The absence of classical separation rules out the possibility of a classical
dynamic program to find the optimum control laws. However, optimal control laws
can still be found in a sequential manner. Based on the two structural results,
we present two sequential methodologies to find optimal control laws. Unlike
classical dynamic programs, each step in our sequential decomposition involves
optimization over a space of functions instead of the space of control actions. 

\subsection {Notation}

Random variables are denoted by upper case letters; their realization by the
corresponding lower case letter. $X_{a:b}$ is a short hand for the vector $(X_a,
X_{a+1}, \dots, X_b)$ while $X^{c:d}$ is a short hand for the vector $(X^c,
X^{c+1}, \dots, X^{d})$. The combined notation $X^{c:d}_{a:b}$ is a short hand
for the vector $(X^j_i : i = a, a+1, \dots, b$, $j = c, c+1, \dots, d)$.
$\PR{\cdot}$ is the probability of an event, $\EXP{\cdot}$ is the expectation of
a random variable. For a collection of functions $\boldsymbol{g}$, we use
$\PR^{\boldsymbol{g}}{\cdot}$ and $\EXP^{\boldsymbol{g}}{\cdot}$ to denote that
the probability measure/expectation depends on the choice of  functions in
$\boldsymbol{g}$ .$\IND_A(\cdot)$ is the indicator function of a set $A$. For
singleton sets $\{a\}$, we also denote $\IND_{\{a\}}(\cdot)$ by $\IND_a(\cdot)$.
For a finite set $A$, $\PSP{A}$ denotes the space of probability mass functions
on $A$. For convenience of exposition, we will assume all sets have finite
cardinality.

\subsection {Model} \label{sec:PF}

Consider a system consisting of a plant and $K$ controllers with decentralized
information. At time $t$, $t=1,\dots,T$, the state of the plant $X_t$ takes
values in $\ALPHABET X$; the control action $U^k_t$ at station $k$,
$k=1,\dots,K$, takes values in $\ALPHABET U^k$. The initial state $X_0$ of the
plant is a random variable. With time, the plant evolves according to
\begin{equation}\label{eq:dynamics}
  X_t = f_t(X_{t-1}, U^{1:K}_t, V_t)
\end{equation}
where $V_t$ is a random variable taking values in $\ALPHABET
V$. $\{V_t; \allowbreak t = 1,\dots, T\}$ is a sequence of independent
random variables that are also independent of $X_0$.

The system has $K$ observation posts. At time $t$, $t=1,\dots, T$, the
observation $Y^k_t$ of post $k$, $k=1,\dots,K$, takes values in $\ALPHABET Y^k$.
These observations are generated according to
\begin{equation}
  \label{eq:obs}
  Y^k_t = h^k_t(X_{t-1}, W^k_t)
\end{equation}
where $W^k_t$ are random variables taking values in $\ALPHABET W^k$. $\{W^k_t;
\allowbreak t =1, \dots, T; \allowbreak k =1,\dots, K\}$ are independent random
variables that are also independent of $X_0$ and $\{V_t; \allowbreak
t=1,\dots,T\}$.

The system has $n$-step delayed sharing. This means that at time $t$, control
station $k$ observes the current observation $Y^k_t$ of observation post $k$,
the $n$ steps old observations $Y^{1:K}_{t-n}$ of all posts, and the $n$ steps
old actions $U^{1:K}_{t-n}$ of all stations. Each station has perfect recall;
so, it remembers everything that it has seen and done in the past. Thus, at time
$t$, data available at station $k$ can be written as $(\Delta_t, \Lambda^k_t)$,
where \[\Delta_t \DEFINED (Y^{1:K}_{1:t-n}, U^{1:K}_{1:t-n})\] is the data known
to all stations and \[\Lambda^k_t \DEFINED (Y^k_{t-n+1:t}, U^k_{t-n+1:t-1})\] is
the additional data known at station $k$, $k=1,\dots,K$. Let \(\mathcal{D}_t\)
be the space of all possible realizations of \(\Delta_t\); and \(\mathcal{L}^k\)
be the space of all possible realizations of \(\Lambda^k_t\). Station $k$
chooses action $U^k_t$ according to a control law $g^k_t$, i.e.,
\begin{equation}
  \label{eq:control}
  U^k_t = g^k_t(\Lambda^k_t, \Delta_t).
\end{equation}

The choice of $\boldsymbol{g} = \{g^k_t; \allowbreak k=1,\dots, K; \allowbreak
t=1,\dots,T\}$ is called a \emph{design} or a \emph{control strategy}.
$\ALPHABET G$ denotes the class of all possible designs. At time $t$, a cost
$c_t(X_t, U^1_t, \dots, U^K_t)$ is incurred. The performance
$\mathcal{J}(\boldsymbol g)$ of a design is given by the expected total cost
under it, i.e.,
\begin{equation} \label{eq:cost}
  \mathcal{J}(\boldsymbol g) 
    = \EXP^{\boldsymbol{g}}{\sum_{t=1}^T c_t(X_t, U^{1:K}_t)}
\end{equation}
where the expectation is with respect to the joint measure on all the system
variables induced by the choice of $\boldsymbol{g}$. We consider the following
problem.
\begin{problem}\label{prob:main}
  Given the statistics of the primitive random variables $X_0$, $\{V_t;
  \allowbreak t=1,\dots,T\}$, $\{W^k_t; \allowbreak k=1,\dots, K; \allowbreak
  t=1,\dots,T\}$, the plant functions $\{f_t; \allowbreak t=1,\dots, T\}$, the
  observation functions $\{h^k_t; \allowbreak k =1,\dots,K; \allowbreak
  t=1,\dots,T\}$, and the cost functions $\{c_t; \allowbreak t=1,\dots,T\}$
  choose a design $\boldsymbol g^*$ from $\ALPHABET G$ that minimizes the
  expected cost given by~\eqref{eq:cost}.
\end{problem}

\subsection {The structural results} \label{sec:results}

Witsenhausen~\cite{Witsenhausen:1971} asserted the following structural result
for Problem~\ref{prob:main}.

\begin{structure}[Witsenhausen~\cite{Witsenhausen:1971})]
  In Problem~\ref{prob:main}, without loss of optimality we can restrict
  attention to control strategies of the form
  \begin{equation} \label{eq:Wit}
    U^k_t = g^k_t(\Lambda^k_t, \PR{X_{t-n} | \Delta_t}).
  \end{equation}
\end{structure}

Witsenhausen's result claims that all control stations can ``compress'' the
common information $\Delta_t$ to a sufficient statistic $\PR{X_{t-n} |
\Delta_t}$. Unlike $\Delta_t$, the size of $\PR{X_{t-n} | \Delta_t}$ does not
increase with time.

As mentioned earlier, Witsenhausen asserted this result without a proof. Varaiya
and Walrand~\cite{WalrandVaraiya:1978} proved that the above separation result
is true for $n=1$ but false for $n>1$. Kurtaran~\cite{Kurtaran:1979} proposed
an alternate structural result for $n>1$.

\begin{structure}[Kurtaran~\cite{Kurtaran:1979}]
  In Problem~\ref{prob:main}, without loss of optimality we can restrict
  attention to control strategies of the form
  \begin{equation}
    U^k_t = g^k_t\big(Y^k_{t-n+1:t}, 
                      \PR^{g^{1:K}_{1:t-1}}{X_{t-n}, U^{1:K}_{t-n+1:t-1} |
                      \Delta_t}\big).
  \end{equation}
\end{structure}

Kurtaran used a different labeling of the time indices, so the
statement of the result in his paper is slightly different from what we have
stated above. 
Kurtaran's result claims that all control stations can ``compress'' the common
information $\Delta_t$ to a sufficient statistic $\PR^{g^{1:K}_{1:t-1}}{X_{t-n},
U^{1:K}_{t-n+1:t-1}|\Delta_t}$, whose size does not increase with time.
  
Kurtaran proved his result for only the terminal time-step and gave an
abbreviated argument for non-terminal time-steps. We believe that his proof is
incomplete for reasons that we will point out in Section~\ref{sec:kurtaran}. In
this paper, we prove two alternative structural results.

\begin{first-structure}[this paper]
  In Problem~\ref{prob:main}, without loss of optimality we can restrict
  attention to control strategies of the form
  \begin{equation} \label{eq:our_result}
    U^k_t = g^k_t\big(\Lambda^k_t, \PR^{g^{1:K}_{1:t-1}}{X_{t-1},
    \Lambda^{1:K}_t | \Delta_t}\big).
  \end{equation}
\end{first-structure}

This result claims that all control stations can ``compress'' the common
information $\Delta_t$ to a sufficient statistic $\PR^{g^{1:K}_{1:t-1}}{X_{t-1},
\Lambda^{1:K}_t | \Delta_t}$, whose size does not increase with time.

\begin{second-structure}[this paper]
  In Problem~\ref{prob:main}, without loss of optimality we can restrict
  attention to control strategies of the form
  \begin{equation} \label{eq:our_result_2}
    U^k_t = g^k_t\big(\Lambda^k_t, \PR{X_{t-n}|\Delta_t}, r^{1:K}_t \big).
  \end{equation}
  where $r^{1:K}_t$ is a collection of partial functions of the previous $n-1$
  control laws of each controller,
  \begin{equation*}
      r^k_t \DEFINED 
      \{(g^k_m(\cdot, Y^k_{m-n+1:t-n}, U^k_{m-n+1:t-n},\Delta_m),
         t-n+1\leq m \leq t-1 \},
  \end{equation*}
  for $k=1,2,\ldots,K$. Observe that $r^k_t$ depends only on the previous $n-1$
  control laws ($g^k_{t-n+1:t-1}$) and the realization of $\Delta_t$ (which
  consists of $Y^{1:K}_{1:t-n},U^{1:K}_{1:t-n}$). This result claims that the
  belief $\PR{X_{t-n}|\Delta_t}$ and the realization of the partial functions
  $r^{1:K}_t$ form a sufficient representation of $\Delta_t$ in order to
  optimally select the control action at time $t$.
\end{second-structure}

Our structural results cannot be derived from Kurtaran's result and vice-versa.
At present, we are not sure of the correctness of Kurtaran's result. As we
mentioned before, we believe that the proof given by Kurtaran is incomplete. We
have not been able to complete Kurtaran's proof; neither have we been able to
find a counterexample to his result. 

Kurtaran's and our structural results differ from those asserted by Witsenhausen
in a fundamental way. The sufficient statistic (also called information state)
$\PR{X_{t-n} | \Delta_t}$ of Witsenhausen's assertion does not depend on the
control strategy. The sufficient statistics 
$\PR^{g^{1:K}_{1:t-1}}{X_{t-n}, U^{1:K}_{t-n+1:t-1}|\Delta_t}$ of Kurtaran's
result and $\PR^{g^{1:K}_{1:t-1}}{X_{t-1}, \Lambda^{1:K}_{t}|\Delta_t}$ of our
first result \emph{depend on the control laws used before time $t$}. Thus, for a
given realization of the primitive random variables till time $t$, the
realization of future information states depend on the choice of control laws at
time $t$. On the other hand, in our second structural result, the belief
$\PR{X_{t-n} | \Delta_t}$ is indeed independent of the control strategy, however
information about the previous $n-1$ control laws is still needed in the form of
the partial functions $r^{1:K}_t$. Since the partial functions $r^{1:K}_t$ do
not depend on control laws used before time $t-n+1$, we conclude that the
information state at time $t$ is separated from the choice of control laws
before time $t-n+1$. We call this a delayed separation between information
states and control laws.

The rest of this paper is organized as follows. We prove our first structural
result in Section~\ref{sec:structural_result}. Then, in
Section~\ref{sec:second_str} we derive our second structural result. We discuss
a special case of delayed sharing information structures in
Section~\ref{sec:aicardi}. We discuss Kurtaran's structural result in
Section~\ref{sec:kurtaran} and conclude in Section~\ref{sec:conclusion}.

\section{Proof of the first structural result} \label{sec:structural_result}

In this section, we prove the structural result~\eqref{eq:our_result} for
optimal strategies of the $K$ control stations. For the ease of notation, we
first prove the result for $K=2$, and then show how to extend it for general
$K$. 
\subsection{Two Controller system ($K=2$)}
The proof for $K=2$ proceeds as follows:
\begin{enumerate}
  \item First, we formulate  a centralized stochastic control problem from the
    point of view of a coordinator who observes the shared information
    \(\Delta_t\), but does not observe the private information $(\Lambda^1_t,
    \Lambda^2_t)$ of the two controllers.

  \item Next, we argue that any strategy for the coordinator's problem can be
    implemented in the original problem and vice versa. Hence, the two problems
    are equivalent.
    
  \item Then, we identify states sufficient for input-output mapping for the
    coordinator's problem.

  \item Finally, we transform the coordinator's problem into a MDP (Markov
    decision process), and obtain a structural result for the coordinator's
    problem. This structural result is also a structural result for the delayed
    sharing information strucutres due to the equivalence between the two
    problems.
\end{enumerate}

Below, we elaborate on each of these stages.

\subsection*{Stage 1}

We consider the following modified problem. In the model described in
Section~\ref{sec:PF}, in addition to the two controllers, a coordinator
 that knows the common (shared) information $\Delta_t$
available to both controllers at time $t$ is present. At time $t$, the coordinator decides the \emph{partial
functions}
\begin{equation*}
  \gamma^k_t : \ALPHABET L^k \mapsto \ALPHABET U^k 
\end{equation*}
for each controller $k$, $k=1,2$. The choice of the partial functions at time
$t$ is based on the realization of the common (shared) information and the
partial functions selected before time $t$. These functions map each
controller's \emph{private information} $\Lambda^k_t$ to its control action
$U^k_t$ at time $t$. The coordinator then informs all controllers of all the
partial functions it selected at time $t$. Each controller then uses its
assigned partial function to generate a control action as follows.
\begin{equation}
  \label{eq:Control1}
  U^k_t = \gamma^k_t(\Lambda^k_t).
\end{equation}

The system dynamics and the cost are same as in the original problem. At next
time step, the coordinator observes the new common observation 
\begin{equation} \label{eq:CoordObs}
  Z_{t+1} \DEFINED 
  \{Y^1_{t-n+1}, Y^2_{t-n+1}, U^1_{t-n+1}, U^2_{t-n+1}\}.
\end{equation}
Thus at the next time, the coordinator knows
$\Delta_{t+1} = Z_{t+1} \cup \Delta_t$ and its choice of all past partial functions and it selects the next partial
functions for each controller. The system proceeds sequentially in this
manner until time horizon $T$.

In the above formulation, the only decision maker is the coordinator: the
individual controllers simply carry out the necessary evaluations prescribed
by~\eqref{eq:Control1}. At time $t$, the coordinator knows the common (shared)
information $\Delta_t$ and all past partial functions $\gamma^{1}_{1:t-1}$ and
$\gamma^{2}_{1:t-1}$. The coordinator uses a decision rule $\psi_t$ to map this
information to its decision, that is,
\begin{gather}
  (\gamma^1_t, \gamma^2_t) 
    = \psi_t(\Delta_t,\gamma^{1}_{1:t-1},\gamma^{2}_{1:t-1}), \\
  \shortintertext{or equivalently,}
  \gamma^k_t = \psi^k_t(\Delta_t, \gamma^{1}_{1:t-1},\gamma^{2}_{1:t-1}), \quad 
  k=1,2.
\end{gather}

The choice of $\boldsymbol\psi = \{\psi_t; \allowbreak t = 1,\dots, T\}$
is called a \emph{coordination strategy}. $\Psi$ denotes the class of
all possible coordination strategies. The performance of a coordinating
strategy is given by the expected total cost under that strategy, that is,
\begin{equation}
  \label{eq:cost-coordinator}
  \hat{\mathcal{J}}(\boldsymbol\psi) = 
  \EXP^{\boldsymbol\psi}{\sum_{t=1}^T c_t(X_t, U^1_t, U^2_t) }
\end{equation}
where the expectation is with respect to the joint measure on all the
system variables induced by the choice of $\boldsymbol\psi$. The coordinator has
to solve the following optimization problem.
 
\begin{problem}[The Coordinator's Optimization Problem]\label{prob:coordinator}
  Given the system model of Problem~\ref{prob:main}, choose a
  coordination strategy $\boldsymbol\psi^*$ from $\Psi$ that minimizes the
  expected cost given by~\eqref{eq:cost-coordinator}.
\end{problem}

\subsection*{Stage 2}

We now show that the Problem~\ref{prob:coordinator} is equivalent to
Problem~\ref{prob:main}. Specifically, we will show that any design
$\boldsymbol{g}$ for Problem~\ref{prob:main} can be implemented by the
coordinator in Problem~\ref{prob:coordinator} with the same value of the problem
objective. Conversely, any coordination strategy $\boldsymbol\psi$ in
Problem~\ref{prob:coordinator} can be implemented in Problem~\ref{prob:main}
with the same value of the performance objective.

Any design $\boldsymbol{g}$ for Problem~\ref{prob:main} can be implemented by
the coordinator in Problem~\ref{prob:coordinator} as follows.  At time $t$ the
coordinator selects partial functions $(\gamma^1_t, \gamma^2_t)$ using the
common (shared) information $\delta_t$ as follows.
\begin{equation} \label{eq:equiv1}
  \gamma^k_t(\cdot) = g^k_t(\cdot, \delta_t) 
  \BYDEFINITION \psi^k_t(\delta_t) ,
  \quad k = 1,2.
\end{equation}

Consider Problems~\ref{prob:main} and~\ref{prob:coordinator}. Use design
$\boldsymbol{g}$ in Problem~\ref{prob:main} and coordination strategy
$\boldsymbol{\psi}$ given by~\eqref{eq:equiv1} in
Problem~\ref{prob:coordinator}. Fix a specific realization of the initial state
$X_0$, the plant disturbance $\{V_t; \allowbreak t=1,\dots,T\}$, and the
observation noise $\{W^1_t,W^2_t; \allowbreak t=1,\dots,T\}$. Then, the choice
of $\boldsymbol{\psi}$ according to~\eqref{eq:equiv1} implies that the
realization of the state $\{X_t; \allowbreak t=1,\dots,T\}$, the observations
$\{Y^1_t, Y^2_t; \allowbreak t=1,\dots,T\}$, and the control actions $\{U^1_t,
U^2_t; \allowbreak t=1,\dots,T\}$ are identical in Problem~\ref{prob:main}
and~\ref{prob:coordinator}. Thus, any design $\boldsymbol{g}$ for
Problem~\ref{prob:main} can be implemented by the coordinator in
Problem~\ref{prob:coordinator} by using a coordination strategy given
by~\eqref{eq:equiv1} and the total expected cost under $\boldsymbol{g}$ in
Problem~\ref{prob:main} is same as the total expected cost under the
coordination strategy given by~\eqref{eq:equiv1} in
Problem~\ref{prob:coordinator}. 

\begin{subequations}\label{eq:equiv2}
  By a similar argument, any coordination strategy $\boldsymbol{\psi}$
  for Problem~\ref{prob:coordinator} can be implemented by the control
  stations in Problem~\ref{prob:main} as follows. At time $1$, both
  stations know $\delta_1$; so, all of them can compute $\gamma^1_1 =
  \psi^1_1(\delta_1)$, $\gamma^2_1 =
  \psi^2_1(\delta_1)$. Then station $k$ chooses action
  $u^k_1 = \gamma^k_1(\lambda^k_1)$. Thus,
  \begin{equation}
    g^k_1(\lambda^k_1, \delta_1) = \psi^k_1(\delta_1)(\lambda^k_1), 
    \quad k = 1,2.
  \end{equation}
  At time $2$, both stations know $\delta_2$ and $\gamma^1_1, \gamma^2_1$, so both of them can
  compute $\gamma^k_2 = \psi^k_2(\delta_2, \gamma^1_1, \gamma^2_1 )$,
  $k=1,2$. Then station $k$ chooses action $u^k_2 =
  \gamma^k_2(\lambda^k_2)$. Thus,
  \begin{equation}
    g^k_2(\lambda^k_2, \delta_2) = \psi^k_2(\delta_2,\gamma^1_1,
    \gamma^2_1)(\lambda^k_2), \quad k = 1,2.
  \end{equation}
  Proceeding this way, at time $t$ both stations know $\delta_t$ and $\gamma^1_{1:t-1}$ and $\gamma^2_{1:t-1}$, so both of them can compute
  $(\gamma^1_{1:t}, \gamma^2_{1:t}) = \psi_t(\delta_t,
    \gamma^1_{1:t-1}, \gamma^2_{1:t-1}) $. 
  Then, station $k$ chooses action $u^k_t = \gamma^k_t(\lambda^k_t)$.
  Thus,
  \begin{equation}
    g^k_t(\lambda^k_t, \delta_t) = \psi^k_t(\delta_t,
    \gamma^1_{1:t-1}, \gamma^2_{1:t-1})(\lambda^k_t),
    \quad k = 1,2.
  \end{equation}
\end{subequations}

Now consider Problems~\ref{prob:coordinator} and~\ref{prob:main}. Use
coordinator strategy $\boldsymbol{\psi}$ in Problem~\ref{prob:coordinator} and
design $\boldsymbol{g}$ given by~\eqref{eq:equiv2} in Problem~\ref{prob:main}.
Fix a specific realization of the initial state $X_0$, the plant disturbance
$\{V_t; \allowbreak t=1,\dots,T\}$, and the observation noise $\{W^1_t, W^2_t;
\allowbreak t=1,\dots,T\}$. Then, the choice of $\boldsymbol{g}$ according
to~\eqref{eq:equiv2} implies that the realization of the state $\{X_t;
\allowbreak t=1,\dots,T\}$, the observations $\{Y^1_t, Y^2_t;\allowbreak
t=1,\dots,T\}$, and the control actions $\{U^1_t, U^2_t; \allowbreak
t=1,\dots,T\}$ are identical in Problem~\ref{prob:coordinator}
and~\ref{prob:main}. Hence, any coordination strategy $\boldsymbol\psi$ for
Problem~\ref{prob:coordinator} can be implemented by the stations in
Problem~\ref{prob:main} by using a design given by~\eqref{eq:equiv2} and the
total expected cost under $\boldsymbol\psi$ in Problem~\ref{prob:coordinator} is
same as the total expected cost under the design given by~\eqref{eq:equiv2} in
Problem~\ref{prob:main}.

Since Problems~\ref{prob:main} and~\ref{prob:coordinator} are equivalent, we
derive structural results for the latter problem. Unlike,
Problem~\ref{prob:main}, where we have multiple control stations, the
coordinator is the only decision maker in Problem~\ref{prob:coordinator}.

\subsection*{Stage 3}     

We now look at Problem~\ref{prob:coordinator} as a controlled input-output
system from the point of view of the coordinator and identify a state sufficient
for input-output mapping. From the coordinator's viewpoint, the input at time
$t$ has two components: a stochastic input that consists of the plant
disturbance $V_t$ and observation noises $W^1_t,W^2_t$; and a controlled input
that consists of the partial functions $\gamma^1_t, \gamma^2_t$. The output is
the observations $Z_{t+1}$ given by~\eqref{eq:CoordObs}. The cost is given by
$c_t(X_t, U^1_t, U^2_t)$. We want to identify a state sufficient for
input-output mapping for this system.

A variable is a state sufficient for input output mapping of a control system if
it satisfies the following properties (see~\cite{Witsenhausen:1976}).
\begin{itemize}
  \item[P1)] The next state is a function of the current state and the
    current inputs.

  \item[P2)] The current output is function of the current state and
    the current inputs.

  \item[P3)] The instantaneous cost is a function of the current state,
    the current control inputs, and the next state.
\end{itemize}

We claim that such a state for Problem~\ref{prob:coordinator} is the
following.

\begin{definition}
  For each $t$ define
  \begin{equation} \label{eq:state}
    S_t \DEFINED (X_{t-1}, \Lambda^1_t, \Lambda^2_t)
  \end{equation}
\end{definition}

Next we show that $S_t$, $t=1,2,\ldots,T+1$, satisfy properties (P1)--(P3).
Specifically, we have the following.

\begin{proposition}\label{prop:state}
  \strut 
  \begin{enumerate}
    \item There exist functions $\hat f_t$, $t=2,\dots,T$ such that
      \begin{equation}
        S_{t+1} = \hat f_{t+1}(S_t, V_t, W^1_{t+1}, W^2_{t+1}, \gamma^1_t,
        \gamma^2_t).
      \end{equation}
    \item There exist functions $\hat h_t$, $t=2,\dots,T$ such that
      \begin{equation}\label{eq:coordinator-observation}
        Z_t = \hat h_t(S_{t-1}).
      \end{equation}
    \item There exist functions $\hat c_t$, $t=1,\dots,T$ such that
      \begin{equation}
        c_t(X_t, U^1_t, U^2_t) = \hat c_t(S_t, \gamma^1_t, \gamma^2_t, S_{t+1}).
      \end{equation}
  \end{enumerate}
\end{proposition}
\begin{proof}
  Part~1 is an immediate consequence of the definitions of $S_t$ and
  $\Lambda^k_t$, the dynamics of the system given by~\eqref{eq:dynamics}, and
  the evaluations carried out by the control stations according
  to~\eqref{eq:Control1}. Part~2 is an immediate consequence of the definitions
  of state $S_t$, observation $Z_t$, and private information $\Lambda^k_t$.
  Part~3 is an immediate consequence of the definition of state and the
  evaluations carried out by the control stations according
  to~\eqref{eq:Control1}.
\end{proof}

\subsection*{Stage 4}
Proposition~\ref{prop:state} establishes $S_t$ as the state sufficient for
input-output mapping for the coordinator's problem. We now define information
states for the coordinator.   
\begin{definition}[Information States]\label{def:info}
For a coordination strategy $\boldsymbol\psi$, define \emph{information states}
$\Pi_t$ as
\begin{equation} \label{eq:define_pi}
  \Pi_t(s_t) \DEFINED 
  \PR^{\boldsymbol\psi}{S_t = s_t | \Delta_t, \gamma^1_{1:t-1},\gamma^2_{1:t-1}}.
\end{equation}
\end{definition}

As shown in Proposition~\ref{prop:state}, the state evolution of $S_t$ depends
on the controlled inputs $(\gamma^1_t, \gamma^2_t)$ and the random noise $(V_t,
W^1_{t+1}, W^2_{t+1})$. This random noise is independent across time.
Consequently, $\Pi_t$ evolves in a controlled Markovian manner as below.
\begin{proposition}\label{prop:info}
  For $t=1,\dots, T-1$, there exists functions $F_t$ (which do not depend on the
  coordinator's strategy) such that
  \begin{equation} \label{eq:info_state_update}
     \Pi_{t+1}  = F_{t+1}(\Pi_{t}, \gamma^1_{t}, \gamma^2_{t}, Z_{t+1}).
  \end{equation}
\end{proposition}
\begin{proof}
  See Appendix~\ref{app:info}.
\end{proof}

At $t=1$, since there is no shared information, $\Pi_1$ is simply the
unconditional probability $\PR{S_1} = \PR{X_0,Y^1_1,Y^2_1}$. Thus, $\Pi_1$ is
fixed a priori from the joint distribution of the primitive random variables and
does not depend on the choice of coordinator's strategy $\psi$.
Proposition~\ref{prop:info} shows that at $t=2,\dots,T$, $\Pi_{t}$ depends on
the strategy $\boldsymbol\psi$ only through the choices of $\gamma^1_{1:t-1}$
and $\gamma^2_{1:t-1}$. Moreover, as shown in Proposition~\ref{prop:state}, the
instantaneous cost at time $t$ can be written in terms of the current and next
states $(S_t, S_{t+1})$ and the control inputs $(\gamma^1_t, \gamma^2_t)$.
Combining the above two properties, we get the following:

\begin{proposition}\label{prop:MDP}
  The process $\Pi_t$, $t=1,2,\ldots,T$ is a controlled Markov chain with
  \(\gamma^1_t,\gamma^2_t\) as the control actions at time $t$, i.e.,
  \begin{equation}
    \PR{\Pi_{t+1}|\Delta_t, \Pi_{1:t}, \gamma^1_{1:t},\gamma^2_{1:t}} =\PR{\Pi_{t+1}|\Pi_{1:t}, \gamma^1_{1:t},\gamma^2_{1:t}} = \PR{\Pi_{t+1}|\Pi_{t},\gamma^1_{t},\gamma^2_{t}}. \label{eq:Markov_State}
  \end{equation}
  Furthermore, there exists a deterministic function $C_t$ such that
  \begin{equation}\label{eq:MDP_Cost}
    \EXP{\hat{c}_t(S_t,\gamma^1_t,\gamma^2_t,S_{t+1})|
      \Delta_t,\Pi_{1:t},\gamma^{1}_{1:t},\gamma^{2}_{1:t}}
    =  C_t(\Pi_t, \gamma^1_1,\gamma^2_t).
  \end{equation}
\end{proposition}
\begin{proof}
  See Appendix~\ref{app:MDP}.
\end{proof}

The controlled Markov property of the process $\{\Pi_t, \allowbreak
t=1,\dots,T\}$ immediately gives rise to the following structural result.

\begin{theorem}\label{thm:coordinator}
  In Problem~\ref{prob:coordinator}, without loss of optimality we can restrict
  attention to coordination strategies of the form
 \begin{equation}
   (\gamma^1_t, \gamma^2_t) = \psi_t(\Pi_t), \quad t = 1, \dots, T.
 \end{equation}
\end{theorem}

\begin{proof}
 From Proposition \ref{prop:MDP}, we conclude that the optimization problem for
 the coordinator is to control the evolution of the controlled Markov process
 $\{\Pi_t$, $t=1,2,\ldots,T\}$ by selecting the partial functions $\{\gamma^1_t,
 \gamma^2_t$, $t=1,2,\ldots,T\}$ in order to minimize $\sum_{t=1}^{T}\EXP
 {C_t(\Pi_t,\gamma^1_t,\gamma^2_t)}$. This is an instance of the well-known
 Markov decision problems where it is known that the optimal strategy is a
 function of the current state. Thus, the structural result follows from Markov
 decision theory~\cite{KumarVaraiya:1986}.
\end{proof}

The above result can also be stated in terms of the original problem.

\begin{theorem}[Structural Result] \label{thm:structural_result}
 In Problem~\ref{prob:main} with $K=2$, without loss of optimality we can
 restrict attention to coordination strategies of the form
 \begin{equation}
   U^k_t = g^k_t(\Lambda^k_t, \Pi_t), \quad k=1,2.
 \end{equation}
 where 
 \begin{equation}
 \Pi_t = \PR^{(g^1_{1:t-1}, g^2_{1:t-1})}
        {X_{t-1}, \Lambda^1_t, \Lambda^2_t | \Delta_t}
 \end{equation}       
 where $\Pi_1 = \PR{X_0,Y^1_1,Y^2_1}$ and for
 $t=2,\ldots,T$, $\Pi_t$ is evaluated as  follows:
 \begin{equation}
  \Pi_{t+1}  = F_{t+1}(\Pi_{t}, g^1_t(\cdot, \Pi_t), g^2_{t}(\cdot, \Pi_t), Z_{t+1})
 \end{equation}
\end{theorem}

\begin{proof}
  Theorem~\ref{thm:coordinator} established the structure of the optimal
  coordination strategy. As we argued in Stage~2, this optimal coordination
  strategy can be implemented in Problem~\ref{prob:main} and is optimal for the
  objective~\eqref{eq:cost}. At $t=1$, $\Pi_1 =
  \PR{X_0,Y^1_1,Y^2_1}$ is known to both controllers and they can use the
  optimal coordination strategy to select partial functions according to:
  \begin{equation*}
    (\gamma^1_1, \gamma^2_1) = \psi_1(\Pi_1) 
  \end{equation*}
 Thus,
 \begin{equation}
  U^k_1 = \gamma^k_1(\Lambda^k_1) 
        = \psi^k_1(\Pi_1)(\Lambda^k_1) 
        \BYDEFINITION g^k_1(\Lambda^k_1,\Pi_1), \quad k=1,2.
 \end{equation}

 At time instant $t+1$, both controllers know $\Pi_t$ and the common
 observations $Z_{t+1} = (Y^1_{t-n+1}, Y^2_{t-n+1}, \allowbreak U^1_{t-n+1},
 U^2_{t-n+1})$; they use the partial functions ($g^1_t(\cdot,\Pi_t),
 g^2_t(\cdot, \Pi_t)$) in equation \eqref{eq:info_state_update} to evaluate
 $\Pi_{t+1}$. The control actions at time $t+1$ are given as:
 \begin{align}
  U^k_{t+1} = \gamma^k_{t+1}(\Lambda^k_{t+1}) 
            &= \psi_{t+1}(\Pi_{t+1})(\Lambda^k_{t+1})  \notag \\
            &\BYDEFINITION g^k_{t+1}(\Lambda^k_{t+1},\Pi_{t+1}),  \label{eq:thm2_eq2}
            \quad k=1,2.
 \end{align}
 Moreover, using the design $\boldsymbol g$ defined according to
 \eqref{eq:thm2_eq2}, the coordinator's information state $\Pi_t$ can also be
 written as:
 \begin{align}
  \Pi_t &=\PR^{\boldsymbol\psi}{X_{t-1}, \Lambda^1_t, \Lambda^2_t | \Delta_t, \gamma^1_{1:t-1},\gamma^2_{1:t-1}} \nonumber \\
  &= \PR^{\boldsymbol g}{X_{t-1}, \Lambda^1_t, \Lambda^2_t |\Delta_t,
      g^{1:2}_1(\cdot,\Pi_1),\ldots,g^{1:2}_{t-1}(\cdot,\Pi_{t-1})} \nonumber \\
  &= \PR^{(g^1_{1:t-1}, g^2_{1:t-1})}{X_{t-1}, \Lambda^1_t, \Lambda^2_t |
      \Delta_t} \label{eq:thm2_eq3}  
 \end{align}
   where we dropped the partial functions from the conditioning terms in
   \eqref{eq:thm2_eq3} because under the given control laws $(g^1_{1:t-1},
   g^2_{1:t-1})$, the partial functions used from time $1$ to $t-1$ can be
   evaluated from $\Delta_t$ (by using Proposition~\ref{prop:info} to evaluate
   $\Pi_{1:t-1}$). 
\end{proof}

Theorem~\ref{thm:structural_result} establishes the first structural result stated in Section~\ref{sec:results} for $K=2$. In the next section, we show how to extend the result for general $K$.
\subsection{Extension to General $K$} \label{sec:extension}
Theorem~\ref{thm:structural_result} for two controllers ($K=2$) can be easily
extended to general $K$ by following the same sequence of arguments as in stages
1 to 4 above. Thus, at time $t$, the  coordinator introduced in Stage~1 now
selects partial functions $\gamma^k_t: \mathcal{L}^k \mapsto \mathcal{U}^k$, for
$k=1,2,\ldots,K$. The state sufficient for input output mapping from the
coordinator's perspective is given as $S_t \DEFINED
(X_{t-1},\Lambda^{1:K}_t)$ and the information state $\Pi_t$ for
the coordinator is
\begin{equation}
 \Pi_t(s_t) \DEFINED 
 \PR^{\boldsymbol\psi}{S_t = s_t | \Delta_t, \gamma^{1:K}_{1:t-1}}.
\end{equation}
Results analogous to Propositions~\ref{prop:state}--\ref{prop:MDP} can now be
used to conclude the structural result of Theorem~\ref{thm:structural_result}
for general~$K$.

\subsection{Sequential Decomposition}
In addition to obtaining the structural result of Theorem~\ref{thm:structural_result},
the coordinator's problem also allows us to write a dynamic program for finding the optimal control 
strategies as shown below. We first focus on the two controller case ($K=2$) and then extend the result to general $K$.

\begin{theorem}\label{thm:seq_decomposition}
 The optimal coordination strategy can be found by the following dynamic program:
 For $t=1,\dots,T$, define the functions $J_{t} : \PSP{S}
  \mapsto \reals$ as follows. For ${\pi} \in \PSP{\ALPHABET S}$ let
  \begin{equation}
    J_{T}(\pi) = \inf _{\tilde\gamma^1,\tilde\gamma^2} 
    \EXP{C_T(\Pi_T,\gamma^1_T,\gamma^2_T) | 
    \Pi_T=\pi,\gamma^1_T=\tilde\gamma^1, \gamma^2_T=\tilde\gamma^2}.
  \end{equation}
  For $t=1,\dots,T-1$, and $\pi \in \PSP{\ALPHABET S}$ let
  \begin{equation}
    J_{t}(\pi) = \inf_{\tilde\gamma^1,\tilde\gamma^2} 
    \EXP{C_t(\Pi_t,\gamma^1_t,\gamma^2_t)+ J_{t+1}(\Pi_{t+1}) |
         \Pi_t=\pi, \gamma^1_t=\tilde\gamma^1, \gamma^2_t=\tilde\gamma^2 }.
  \end{equation}
  The arg inf $(\gamma^{*,1}_t, \gamma^{*,2}_t)$ in the RHS of $J_t(\pi)$
  is the optimal action for the coordinator at time $t$ then $\Pi_t = \pi$.
  Thus,
  \begin{equation*}
    (\gamma^{*,1}_t, \gamma^{*,2}_t) = \phi^*_t(\pi_t)
  \end{equation*}
  The corresponding control strategy for Problem~\ref{prob:main}, given
  by~\eqref{eq:equiv2} is optimal for Problem~\ref{prob:main}. 
\end{theorem}
\begin{proof}
  As in Theorem~\ref{thm:coordinator}, we use the fact that the coordinator's
  optimization problem can be viewed as a Markov decision problem with $\Pi_t$
  as the state of the Markov process. The dynamic program follows from  standard
  results in Markov decision theory~\cite{KumarVaraiya:1986}. The optimality of
  the corresponding control strategy for Problem~\ref{prob:main} follows from
  the equivalence between the two problems.
\end{proof}

The dynamic program of Theorem~\ref{thm:seq_decomposition} can be extended to
general $K$ in a manned similar to Section~\ref{sec:extension}

\subsection{Computational Aspects}

In the dynamic program for the coordinator in
Theorem~\ref{thm:seq_decomposition}, the value functions at each time are
functions defined on the continuous space $\PSP{\ALPHABET S}$, whereas the
minimization at each time step is over the finite set of functions from the
space of realizations of the private information of controllers
($\mathcal{L}^k$, $k=1,2$) to the space of control actions ($\mathcal{U}^k$,
$k=1,2$). While dynamic programs with continuous state space can be hard to
solve, we note that our dynamic program resembles the dynamic program for
partially observable Markov decision problems (POMDP). In particular, just as in
POMDP, the value-function at time $T$ is piecewise linear in $\Pi_{T}$ and by
standard backward recursion, it can be shown that value-function at time $t$ is
piecewise linear and concave function of $\Pi_t$. (See
Appendix~\ref{app:convex}). Indeed, the coordinator's problem can be viewed as a
POMDP, with $S_t$ as the underlying partially observed state and the belief
$\Pi_t$ as the information state of the POMDP. The characterization of value
functions as piecewise linear and concave is utilized to find computationally
efficient algorithms for POMDPs. Such algorithmic solutions to general POMDPs
are well-studied and can be employed here. We refer the reader to
\cite{Zhang:2009} and references therein for a review of algorithms to solve
POMDPs.

\subsection{One-step Delay}

We now focus on the one-step delayed sharing information structure, i.e., when
$n=1$. For this case, the structural result~\eqref{eq:Wit} asserted by
Witsenhausen is correct~\cite{WalrandVaraiya:1978}. At first glance, that
structural result looks different from our structural
result~\eqref{eq:our_result} for $n=1$. In this section, we show that for $n=1$,
these two structural results are equivalent. 

As before, we consider the two-controller system ($K=2$). When delay $n=1$, we
have
\begin{gather*}
  \Delta_t = (Y^{1}_{1:t-1},Y^{2}_{1:t-1},U^{1}_{1:t-1},U^{2}_{1:t-1}), \\
  \Lambda^1_t = (Y^1_t), \quad \Lambda^2_t = (Y^2_t), \\
  \shortintertext{and}
  Z_{t+1} = (Y^1_t,Y^2_t,U^1_t,U^2_t). 
\end{gather*}

The result of Theorem~\ref{thm:structural_result} can now be restated for this
case as follows:
\begin{corollary}
  In Problem~\ref{prob:main} with $K=2$ and $n=1$, without loss of optimality we
  can restrict attention to control strategies of the form:
  \begin{equation} \label{eq:one_step_result_1}
    U^k_t = g^k_t(Y^k_t, \Pi_t), \quad k=1,2.
  \end{equation}
  where 
  \begin{equation} \label{eq:first_one_step_result}
  \Pi_t \DEFINED 
  \PR^{(g^1_{1:t-1}, g^2_{1:t-1})}{X_{t-1}, Y^1_t, Y^2_t| \Delta_t} 
\end{equation}
\end{corollary} 

We can now compare our result for one-step delay with the structural
result~\eqref{eq:Wit}, asserted in~\cite{Witsenhausen:1971} and 
proved in \cite{WalrandVaraiya:1978}. For $n=1$, this result state that
without loss of optimality, we can restrict attention to control laws of the
form:
\begin{equation} \label{eq:WV}
  U^k_t = g^k_t(Y^k_t, \PR{X_{t-1}|\Delta_t}), \quad k = 1,2.
\end{equation}

The above structural result can be recovered
from~\eqref{eq:first_one_step_result} by observing that there is a one-to-one
correspondence between $\Pi_t$ and the belief $\PR{X_{t-1}|\Delta_t}$. We first
note that 
\begin{align}
\Pi_t &= \PR^{(g^1_{1:t-1}, g^2_{1:t-1})}{X_{t-1}, Y^1_t, Y^2_t| \Delta_t} \notag \\
      &= \PR{Y^1_t|X_{t-1}}\cdot\PR{Y^2_t|X_{t-1}}
      \cdot
      \PR^{(g^1_{1:t-1}, g^2_{1:t-1})}{X_{t-1}| \Delta_t}
\end{align}
As pointed out in~\cite{Witsenhausen:1971, WalrandVaraiya:1978} (and
proved later in this paper in Proposition~\ref{prop:equiv_info}), the last
probability does not depend on the functions $(g^1_{1:t-1}, g^2_{1:t-1})$.
Therefore,
\begin{equation} \label{eq:one_step_eqn}
\Pi_t = \PR{Y^1_t|X_{t-1}}\cdot\PR{Y^2_t|X_{t-1}}\cdot\PR{X_{t-1}| \Delta_t}
\end{equation}
Clearly, the belief $\PR{X_{t-1}|\Delta_t}$ is a marginal of $\Pi_t$ and
therefore can be evaluated from $\Pi_t$. Moreover, given the belief
$\PR{X_{t-1}|\Delta_t}$, one can evaluate $\Pi_t$ using equation
\eqref{eq:one_step_eqn}. This one-to-one correspondence between $\Pi_t$ and
$\PR{X_{t-1} | \Delta_t}$ means that the structural result proposed in this
paper for $n=1$ is effectively equivalent to the one proved
in~\cite{WalrandVaraiya:1978}.

\section{Proof of the second structural result} \label{sec:second_str}

In this section we prove the second structural result~\eqref{eq:our_result_2}.
As in Section~\ref{sec:structural_result}, we prove the result for $K=2$ and
then show how to extend it for general $K$. To prove the result, we reconsider
the coordinator's problem at Stage~3 of Section~\ref{sec:structural_result} and
present an alternative characterization for the coordinator's optimal strategy
in Problem~\ref{prob:coordinator}. The main idea in this section is to use the
dynamics of the system evolution and the observation equations (equations
\eqref{eq:dynamics} and \eqref{eq:obs}) to find an equivalent representation of
the coordinator's information state.
We also contrast this information state with that proposed by Witsenhausen.

\subsection{Two controller system ($K=2$)}

Consider the coordinator's problem with $K=2$. Recall that $\gamma^1_t$ and
$\gamma^2_t$ are the coordinator's actions at time $t$. $\gamma^k_t$ maps the
private information of the $k^{th}$ controller ($Y^k_{t-n+1:t},U^k_{t-n+1:t-1}$)
to its action $U^k_t$. In order to find an alternate characterization of
coordinator's optimal strategy, we need the following definitions: 
\begin{definition}\label{def:equiv_info}
  For a coordination strategy $\boldsymbol\psi$, and for $t=1,2,\ldots,T$ we
  define the following:
  \begin{enumerate}
    \item $\Theta_t \DEFINED \PR{X_{t-n}|\Delta_t}$

    \item For $k=1,2$, define the following partial functions of $\gamma^k_m$
      \begin{equation}
        r^k_{m,t}(\cdot) \DEFINED 
        \gamma^k_m(\cdot, Y^{k}_{m-n+1:t-n}, U^{k}_{m-n+1:t-n}), 
            \quad m= t-n+1,t-n+2, \ldots, t-1 \label{eq:define_r_1}
       \end{equation} 
       Since $\gamma^k_m$ is a function that maps
       ($Y^k_{m-n+1:m},U^k_{m-n+1:m-1}$) to  $U^k_m$, $r^k_{m,t}(\cdot)$ is a
       function that maps ($Y^k_{t-n+1:m},U^k_{t-n+1:m-1}$) to  $U^k_m$. We
       define a collection of these partial functions as follows:
       \begin{equation}
         r^{k}_t \DEFINED (r^k_{m,t}, m=t-n+1,t-n+2,\ldots,t-1) 
         \label{eq:define_r_2}
       \end{equation}
       Note that for $n=1$, $r^k_t$ is empty.
   \end{enumerate}
\end{definition}

We need the following results to address the coordinator's problem:
\begin{proposition}\label{prop:equiv_info}
   \begin{enumerate}
   \item For $t=1,\dots, T-1$, there exists functions $Q_t,Q^k_t$, $k=1,2$,
     (which do not depend on the coordinator's strategy) such that
     \begin{align} \label{eq:info_state_update_2}
       \Theta_{t+1} & = Q_t(\Theta_t,Z_{t+1}) \notag\\
        r^{k}_{t+1} &= Q^k_t(r^{k}_t,Z_{t+1},\gamma^k_t)
     \end{align}           
  
   \item The coordinator's information state $\Pi_t$ is a function of
     $(\Theta_t,r^1_t,r^2_t)$. Consequently, for $t=1,\dots, T$, there exist
     functions $\hat{C}_t$ (which do not depend on the coordinator's strategy)
     such that
     \begin{equation} \label{eq:info_state_update_3}
       \EXP{\hat{c}_t(S_t,\gamma^1_t,\gamma^2_t,S_{t+1})|
            \Delta_t,\Pi_{1:t},\gamma^{1}_{1:t},\gamma^{2}_{1:t}}
        =  \hat{C}_t(\Theta_t,r^1_t,r^2_t,\gamma^1_t,\gamma^2_t)
      \end{equation}    
      
   \item The process $(\Theta_t,r^1_t,r^2_t)$, $t=1,2,\ldots,T$ is a controlled
     Markov chain with \(\gamma^1_t,\gamma^2_t\) as the control actions at time
     $t$, i.e.,
     \begin{align}
        &\PR{\Theta_{t+1},r^1_{t+1},r^2_{t+1}|
        \Delta_t,\Theta_{1:t},r^1_{1:t},r^2_{1:t},
        \gamma^1_{1:t},\gamma^2_{1:t}} \notag \\
        &\quad=
        \PR{\Theta_{t+1},r^1_{t+1},r^2_{t+1}|\Theta_{1:t},r^1_{1:t},r^2_{1:t},
        \gamma^1_{1:t},\gamma^2_{1:t}} \notag \\
        &\quad= 
        \PR{\Theta_{t+1},r^1_{t+1},r^2_{t+1}|
          \Theta_{t},r^1_{t},r^2_{t},\gamma^1_{t},\gamma^2_{t}}. 
          \label{eq:Markov_State_2}
      \end{align}
   \end{enumerate}        
\end{proposition}

\begin{proof}
  See Appendix~\ref{proof:equiv_info}.
\end{proof} 

At $t=1$, since there is no sharing of information, $\Theta_1$ is simply the
unconditioned probability $\PR{X_0}$. Thus, $\Theta_1$ is fixed a priori from
the joint distribution of the primitive random variables and does not depend on
the choice of the coordinator's strategy $\psi$. Proposition 4 shows that the
update of $\Theta_t$ depends only on $Z_{t+1}$ and not on the coordinator's
strategy. Consequently, the belief $\Theta_t$ depends only on the distribution
of the primitive random variables and the realizations of $Z_{1:t}$. We can now
show that the coordinator's optimization problem can be viewed as an MDP with
$(\Theta_t,r^1_t,r^2_t)$, $t=1,2,\ldots,T$ as the underlying Markov process.


\begin{theorem} \label{thm:equiv_info}
  $(\Theta_t,r^1_t,r^2_t)$ is an information state for the coordinator. That is,
  there is an optimal coordination strategy of the form:
  \begin{equation}
   (\gamma^1_t, \gamma^2_t) = 
      \psi_t(\Theta_t,r^1_t,r^2_t), \quad t = 1, \dots, T.
  \end{equation}
  Moreover, this optimal coordination strategy can be found by the following
  dynamic program:
 \begin{equation}
    J_{T}(\theta, \tilde{r}^1,\tilde{r}^2) = 
    \inf _{\tilde\gamma^1,\tilde\gamma^2}
    \EXP{\hat{C}_T(\Theta_T,r^1_T,r^2_T,\gamma^1_T,\gamma^2_T)|
     \Theta_T=\theta, r^1_T=\tilde{r}^1,r^2_T=\tilde{r}^2,
    \gamma^1_T=\tilde\gamma^1,\gamma^2_T=\tilde\gamma^2}.
  \end{equation}
  For $t=1,\dots,T-1$, let
  \begin{equation} \label{eq:equiv_info_dp}
    J_{t}(\theta, \tilde{r}^1,\tilde{r}^2) = 
    \inf_{\tilde\gamma^1,\tilde\gamma^2}
    \EXP{\hat{C}_t(\Theta_t,r^1_t,r^2_t,\gamma^1_1,\gamma^2_t) + 
        J_{t+1}(\Theta_{t+1}, r^1_{t+1},r^2_{t+1}) |
    \Theta_t,=\theta, 
    \begin{array}{l}
      r^1_t=\tilde{r}^1,r^2_t=\tilde{r}^2,\\
      \gamma^1_t=\tilde\gamma^1,\gamma^2_t=\tilde\gamma^2 
    \end{array}}.
  \end{equation}
  where $\theta \in \PSP{\mathcal{X}}$, and $\tilde{r}^1,\tilde{r}^2$ are
  realizations of partial functions defined in~\eqref{eq:define_r_1}
  and~\eqref{eq:define_r_2}. The arg inf $(\gamma^{*,1}_t, \gamma^{*,2}_t)$ in
  the RHS of \eqref{eq:equiv_info_dp} is the optimal action for the coordinator
  at time $t$ when $(\Theta_t,r^1_t,r^2_t) = (\theta, \tilde{r}^1,\tilde{r}^2)$.
  Thus,
  \begin{equation*}
    (\gamma^{*,1}_t, \gamma^{*,2}_t) = \psi^*_t(\Theta_t,r^1_t,r^2_t)
  \end{equation*}
  The corresponding control strategy for Problem~\ref{prob:main}, given
  by~\eqref{eq:equiv2} is optimal for Problem~\ref{prob:main}. 
\end{theorem} 

\begin{proof}
  Proposition~\ref{prop:equiv_info} implies that the coordinator's optimization
  problem can be viewed as an MDP with $(\Theta_t,r^1_t,r^2_t)$,
  $t=1,2,\ldots,T$ as the underlying Markov process and
  $\hat{C}_t(\Theta_t,r^1_t,r^2_t,\gamma^1_t,\gamma^2_t)$ as the instantaneous
  cost. The MDP formulation implies the result of the theorem.
\end{proof}

The following result follows from Theorem~\ref{thm:equiv_info}.

\begin{theorem}[Second Structural Result] \label{thm:second_structural_result}
  In Problem~\ref{prob:main} with $K=2$, without loss of optimality we can
  restrict attention to coordination strategies of the form
  \begin{equation}
    U^k_t = g^k_t(\Lambda^k_t, \Theta_t,r^1_t,r^2_t), \quad k=1,2.
  \end{equation}
  where 
  \begin{equation}
    \Theta_t = \PR{X_{t-n}| \Delta_t}
  \end{equation}      
  and
  \begin{equation}
    r^k_t = \{(g^k_m(\cdot, Y^k_{m-n+1:t-n}, U^k_{m-n+1:t-n},\Delta_m), t-n+1\leq m \leq t-1 \}
  \end{equation}
\end{theorem}

\begin{proof}
  As in Theorem~\ref{thm:structural_result}, equations~\eqref{eq:equiv2} can be
  used to identify an optimal control strategy for each controller from the
  optimal coordination strategy given in Theorem~\ref{thm:equiv_info}.  
\end{proof} 

Theorem~\ref{thm:equiv_info} and Theorem~\ref{thm:second_structural_result} can
be easily extended for $K$ controllers by identifying $(\Theta_t,r^{1:K}_t)$ as
the information state for the coordinator.

\subsection{Comparison to Witsenhausen's Result}

We now compare the result of Theorem~\ref{thm:equiv_info} to Witsenhausen's
conjecture which states that there exist optimal control strategies of the form:
\begin{equation} \label{eq:Wit_2}
  U^k_t = g^k_t(\Lambda^k_t, \PR{X_{t-n} | \Delta_t}).
\end{equation}
Recall that Witsenhausen's conjecture is true for $n=1$ but false for $n>1$.
Therefore, we consider the cases $n=1$ and $n>1$ separately:

\subsubsection*{Delay $n=1$}

For a two-controller system with $n=1$, we have
\begin{gather*}
  \Delta_t = (Y^{1}_{1:t-1},Y^{2}_{1:t-1},U^{1}_{1:t-1},U^{2}_{1:t-1}), \\
  \Lambda^1_t = (Y^1_t), \quad \Lambda^2_t = (Y^2_t), \\
  \shortintertext{and}
  r^1_t = \emptyset, \quad r^2_t = \emptyset
\end{gather*}
Therefore, for $n=1$, Theorem~\ref{thm:second_structural_result} implies that
there exist optimal control strategies of the form:
\begin{equation} 
  U^k_t = g^k_t(\Lambda^k_t, \PR{X_{t-n}|\Delta_t}), \quad k=1,2. 
  \label{eq:one_step_result}
\end{equation}
Equation \eqref{eq:one_step_result} is the same as equation~\eqref{eq:Wit_2} for
$n=1$. Thus, for $n=1$, the result of Theorem~\ref{thm:equiv_info} coincides
with Witsenhausen's conjecture which was proved in~\cite{WalrandVaraiya:1978}.  

\subsubsection*{Delay $n>1$}

Witsenhausen's conjecture implied that the controller $k$ at time $t$ can choose
its action based only on the knowledge of $\Lambda^k_t$ and
$\PR{X_{t-n}|\Delta_t}$, without any dependence on the choice of previous
control laws ($g^{1:2}_{1:t-1}$). In other words, the argument of the control
law $g^k_t$ (that is, the information state at time $t$) is separated from
$g^{1:2}_{1:t-1}$. However, as Theorem~\ref{thm:second_structural_result} shows,
such a separation is not true because of the presence of the collection of
partial functions $r^1_t,r^2_t$ in the argument of the optimal control law at
time $t$. These partial functions depend on the choice of previous $n-1$ control
laws. Thus, the argument of $g^k_t$ depends on the choice of
$g^{1:2}_{t-n+1:t-1}$. One may argue that
Theorem~\ref{thm:second_structural_result} can be viewed as a \emph{delayed or
partial} separation since  the information state for the control law $g^k_t$ is
separated from the choice of control laws before time $t-n+1$.

Witsenhausen's conjecture implied that controllers employ common information
only to form a belief on the state $X_{t-n}$; the controllers do not need to use
the common information to guess each other's behavior from $t-n+1$ to the
current time $t$. Our result disproves this statement. We show that in addition
to forming the belief on $X_{t-n}$, each agent should use the common information
to predict the actions of other agents by means of the partial functions
$r^1_t,r^2_t$.  
   

\section{A Special Case of Delayed Sharing Information Structure} 
\label{sec:aicardi}

Many decentralized systems consist of coupled subsystems, where each subsystem
has a controller that perfectly observes the state of the subsystem. If all
controllers can exchange their observations and actions with a delay of $n$
steps, then the system is a special case of the $n$-step delayed sharing
information structure with the following assumptions:
\begin{enumerate}
    \item \emph{Assumption 1:} At time $t=1,\dots,T$, the state of the system is
      given as the vector $X_t \DEFINED (X_t^{1:K})$, where $X^i_t$ is the state
      of subsystem $i$. 

    \item \emph{Assumption 2:} The observation equation of the $k^{th}$
      controller is given as: 
    \begin{equation}
      Y^k_t = X^k_t
    \end{equation}
\end{enumerate}

This model is the same as the model considered in \cite{Aicardi:1987}. Clearly,
the first structural result and the sequential decomposition of
Section~\ref{sec:structural_result} apply here as well with the observations
$Y^k_t$ being replaced by $X^k_t$. Our second structural result simplifies when
specialized to this model. Observe that in this model
\begin{align}
  \Delta_t = (Y^{1:K}_{1:t-n},U^{1:K}_{1:t-n}) = (X_{1:t-n},U^{1:K}_{1:t-n})
\end{align}
and therefore the belief,
\begin{align} 
  \Theta_t = \PR{X_{t-n}|\Delta_t}
\end{align}
is $1$ for the true realization of $X_{t-n}$ and $0$ otherwise. The result of
Theorem~\ref{thm:equiv_info} can now be restated for this case as follows:
\begin{corollary}
  In Problem~\ref{prob:main} with assumptions~1 and~2, there is an optimal
  coordination strategy of the form:
  \begin{equation}
    (\gamma^1_t, \gamma^2_t) = \psi_t(X_{t-n},r^1_t,r^2_t), 
    \quad t = 1, \dots, T.
  \end{equation}
  Moreover, this optimal coordination strategy can be found by the following
  dynamic program:
  \begin{equation}
    J_{T}(x, \tilde{r}^1,\tilde{r}^2) = 
    \inf _{\tilde\gamma^1,\tilde\gamma^2} 
    \EXP{\hat{C}_T(X_{T-n},r^1_T,r^2_T,\gamma^1_T,\gamma^2_T)|
    X_{T-n}=x, r^1_T=\tilde{r}^1,r^2_T=\tilde{r}^2,
    \gamma^1_T=\tilde\gamma^1,\gamma^2_T=\tilde\gamma^2}.
  \end{equation}
  For $t=1,\dots,T-1$, let
  \begin{equation}
    J_{t}(x, \tilde{r}^1,\tilde{r}^2) = 
    \inf_{\tilde\gamma^1,\tilde\gamma^2} 
    \EXP{\hat{C}_t(X_{t-n},r^1_t,r^2_t,\gamma^1_1,\gamma^2_t) + 
    J_{t+1}(X_{t-n+1}, r^1_{t+1},r^2_{t+1}) |
    \begin{array}{l}
      X_{t-n}=x,\\r^1_t=\tilde{r}^1,r^2_t=\tilde{r}^2,\\
      \gamma^1_t=\tilde\gamma^1,\gamma^2_t=\tilde\gamma^2
    \end{array}}.
  \end{equation}
\end{corollary} 
We note that the structural result and the sequential decomposition in the
corollary above is analogous to Theorem 1 of \cite{Aicardi:1987}. 

\section{Kurtaran's Separation Result} \label{sec:kurtaran}

In this section, we focus on the structural result proposed by
Kurtaran~\cite{Kurtaran:1979}. We restrict to the two controller system ($K=2$)
and delay $n=2$. For this case, we have
\begin{gather*}
  \Delta_t = (Y^{1}_{1:t-2},Y^{2}_{1:t-2},U^{1}_{1:t-2},U^{2}_{1:t-2}), \\
  \Lambda^1_t = (Y^1_t, Y^1_{t-1}, U^1_{t-1}), \quad \Lambda^2_t = (Y^2_t,
  Y^2_{t-1}, U^2_{t-1}), \\
  \shortintertext{and}
  Z_{t+1} = (Y^1_{t-1},Y^2_{t-1},U^1_{t-1},U^2_{t-1}). 
\end{gather*}

Kurtaran's structural result for this case states that without loss of
optimality we can restrict attention to control strategies of the form:
\begin{equation}
  U^k_t = g^k_t(\Lambda^k_t,\Phi_t),  \quad k=1,2,
\end{equation}
where 
\[ \Phi_t \DEFINED \PR^{\boldsymbol g}{X_{t-2}, U^1_{t-1},U^2_{t-1}|\Delta_t}.
\]
Kurtaran~\cite{Kurtaran:1979} proved this result for the terminal time-step~$T$
and simply stated that the result for $t=1,\dots,T-1$ can be established by the
dynamic programming argument given in~\cite{Kurtaran:1976}. We believe that this
is not the case.

In the dynamic programming argument in~\cite{Kurtaran:1976}, a critical step is
the update of the information state $\Phi_t$, which is given
by~\cite[Eq~(30)]{Kurtaran:1976}. For the result presented
in~\cite{Kurtaran:1979}, the corresponding equation is
\begin{equation}\label{eq:wrong_update}
 \Phi_{t+1} = F_t(\Phi_t,Y^1_{t-1},Y^2_{t-1},U^1_{t-1},U^2_{t-1}).
\end{equation}
We believe that such an update equation cannot be established.

To see the difficulty in establishing~\eqref{eq:wrong_update}, lets follow an
argument similar to the proof of~\cite[Eq~(30)]{Kurtaran:1976} given
in~\cite[Appendix~B]{Kurtaran:1976}. For a fixed strategy $\boldsymbol{g}$, and
a realization $\delta_{t+1}$ of $\Delta_{t+1}$, the realization $\varphi_{t+1}$
of $\Phi_{t+1}$ is given by
\begin{align}
  \varphi_{t+1} &= \PR{x_{t-1}, u^1_{t},u^2_{t}|\delta_{t+1}} \notag \\
                &= \PR{x_{t-1},u^1_{t},u^2_{t}|\delta_t, y^1_{t-1},y^2_{t-1},u^1_{t-1},u^2_{t-1}} \notag \\
                &= \frac {\PR{x_{t-1},u^1_{t},u^2_{t},y^1_{t-1},y^2_{t-1},u^1_{t-1},u^2_{t-1}|\delta_t}}
                         {
                         {
                         {\sum\limits_{(x',a^1, a^2) \in \ALPHABET X \times
                         \ALPHABET U^1 \times \ALPHABET U^2}}
                         \vphantom{\sum\limits^{-}}
                          \PPR( X_{t-1} = x', U^1_t = a^1, U^2_t = a^2,}
                                     {y^1_{t-1}, y^2_{t-1}, u^1_{t-1}, u^2_{t-1} \mid \delta_t)}}
\end{align}
The numerator can be expressed as:
\begin{align}
  \hskip 2em & \hskip -2em
 \PR{x_{t-1},u^1_{t},u^2_{t},y^1_{t-1},y^2_{t-1},u^1_{t-1},u^2_{t-1}|\delta_t} \notag \\
 \displaybreak[2]
 &= \smashoperator[l]{\sum_{(x_{t-2},y^1_t,y^2_t) \in \ALPHABET X \times
 \ALPHABET Y^1 \times \ALPHABET  Y^2}} 
      \Pr(x_{t-1},u^1_{t},u^2_{t},y^1_{t-1},y^2_{t-1},
          u^1_{t-1},u^2_{t-1},x_{t-2},y^1_t,y^2_t|\delta_t)
    \notag \\
 &= \smashoperator[l]{\sum_{(x_{t-2},y^1_t,y^2_t) \in \ALPHABET X \times
 \ALPHABET Y^1 \times \ALPHABET  Y^2}} 
 \IND_{g^1_t(\delta_t,u^1_{t-1},y^1_{t-1},y^1_t)}[u^1_t] 
 \cdot
 \IND_{g^2_t(\delta_t,u^2_{t-1},y^2_{t-1},y^2_t)}[u^2_t] 
 \cdot   
 \PR{y^1_t|x_{t-1}} \cdot \PR{y^2_t|x_{t-1}}\notag \\ 
 & \quad \cdot \PR{x_{t-1}|x_{t-2},u^1_{t-1},u^2_{t-1}} 
 \cdot
 \IND_{g^1_{t-1}(\delta_{t-1},u^1_{t-2},y^1_{t-2},y^1_{t-1})}[u^1_{t-1}] 
 \cdot
 \IND_{g^2_t(\delta_{t-1},u^2_{t-2},y^2_{t-2},y^2_{t-1})}[u^2_{t-2}] \notag
 \\
 & \quad \cdot \PR{y^1_{t-1}|x_{t-2}}\cdot \PR{y^2_{t-1}|x_{t-2}}
 \cdot
 \PR{x_{t-2}|\delta_t} \label{eq:kurtaran_update}
\end{align}
If, in addition to $\varphi_t$, $y^1_{t-1}$, $y^2_{t-1}$, $u^1_{t-1}$, and
$u^2_{t-1}$, each term of~\eqref{eq:kurtaran_update} depended only on terms that
are being summed over ($x_{t-2}$, $y^1_{t}$, $y^2_{t}$),
then~\eqref{eq:kurtaran_update} would prove~\eqref{eq:wrong_update}. However,
this is not the case: the first two terms also depend on $\delta_t$. Therefore,
the above calculation shows that $\varphi_{t+1}$ is a function of
$\varphi_{t},Y^1_{t-1},Y^2_{t-1},U^1_{t-1},U^2_{t-1}$ \emph{and} $\delta_t$.
This dependence on $\delta_t$ is not an artifact of the order in which we
decided to use the chain rule in~\eqref{eq:kurtaran_update} (we choose the
natural sequential order in the system). No matter how we try to write
$\varphi_{t+1}$ in terms of $\varphi_t$, there will be a dependence on
$\delta_t$. 

The above argument shows that it is not possible to
establish~\eqref{eq:wrong_update}. Consequently, the dynamic programming
argument presented in~\cite{Kurtaran:1976} breaks down when working with the
information state of~\cite{Kurtaran:1979}, and, hence, the proof
in~\cite{Kurtaran:1979} is incomplete. So far, we have not been able to correct
the proof or find a counterexample to it.

\section {Conclusion} \label{sec:conclusion}

We studied the stochastic control problem with $n$-step delay sharing
information structure and established two structural results for it. Both the
results characterize optimal control laws with time-invariant domains. Our
second result also establishes a partial separation result, that is, it shows
that the information state at time $t$, is separated from choice of laws before
time $t-n+1$. Both the results agree with Witsenhausen's conjecture for $n=1$.
To derive our structural results, we formulated an alternative problem from the
point of a coordinator of the system. We believe that this idea of formulating
an alternative problem from the point of view of a coordinator which has access
to information common to all controllers is also useful for general
decentralized control problems, as is illustrated
by~\cite{NayyarTeneketzis:2008} and~\cite{MahajanNayyarTeneketzis:2008}.

\appendix

\section{Proof of Proposition~\ref{prop:info}}  \label{app:info}

Fix a coordinator strategy $\boldsymbol\psi$. Consider a realization
$\delta_{t+1}$ of the common information $\Delta_{t+1}$. Let $(\tilde
\gamma^1_{1:t}, \tilde \gamma^2_{1:t})$ be the corresponding choice of partial
functions until time $t$. Then, the realization $\pi_{t+1}$ of $\Pi_{t+1}$ is
given by
\begin{equation}\label{eq:app-info-1}
  \pi_{t+1}(s_{t+1}) = \PR^{\boldsymbol\psi}{S_{t+1} = s_{t+1} | \delta_{t+1}, 
                       \tilde \gamma^1_{1:t}, \tilde \gamma^2_{1:t}}.
\end{equation}
Using Proposition~\ref{prop:state} , this can be written as
\begin{align} 
  \hskip 2em & \hskip -2em
  \sum_{s_t, v_t, w^1_{t+1}, w^2_{t+1}} 
  \IND_{s_{t+1}}( \hat f_{t+1}(s_t, v_t, w^1_{t+1}, w^2_{t+1}, \tilde \gamma^1_t, \tilde \gamma^2_t))
   \cdot 
   \PR{V_t = v_t} \cdot\PR{W^1_{t+1} = w^1_{t+1}}
   \notag \\
   &\quad \cdot
   \PR{W^2_{t+1} = w^2_{t+1}}
   \cdot 
   \PR^{\boldsymbol\psi}{S_t = s_t | \delta_{t+1}, \tilde \gamma^1_{1:t}, \tilde \gamma^2_{1:t}}.
\label{eq:app-info-2}
\end{align}
Since $\delta_{t+1} = (\delta_t, z_{t+1})$, the last term
of~\eqref{eq:app-info-2} can be written as
\begin{equation}
  \PR^{\boldsymbol\psi}{S_t = s_t | \delta_{t}, z_{t+1}, \tilde \gamma^1_{1:t}, \tilde
  \gamma^2_{1:t}} 
  = \frac
  {\PR^{\boldsymbol\psi}{S_t = s_t, Z_{t+1} = z_{t+1} | \delta_t, \tilde \gamma^1_{1:t}, \tilde \gamma^2_{1:t}}}
  {\sum_{s'} \PR^{\boldsymbol\psi}{S_t = s', Z_{t+1} = z_{t+1} | \delta_t, \tilde \gamma^1_{1:t}, \tilde \gamma^2_{1:t}}}.
  \label{eq:app-info-3}
\end{equation}

We can use~\eqref{eq:coordinator-observation} and the sequential order in which
the system variables are generated to write
\begin{align}
  \hskip 2em & \hskip -2em
  \PR^{\boldsymbol\psi}{S_t = s_t, Z_{t+1} = z_{t+1} | \delta_t, \tilde \gamma^1_{1:t}, \tilde \gamma^2_{1:t}} 
  \notag \\
  &= \IND_{\hat h_t(s_t)}(z_{t+1}) 
     \cdot
     \PR^{\boldsymbol\psi}{S_t = s_t | \delta_t, 
                  \tilde \gamma^1_{1:t-1}, \tilde \gamma^2_{1:t-1}} \notag \\
  &= \IND_{\hat h_t(s_t)}(z_{t+1}) \cdot\pi_t(s_t).
  \label{eq:app-info-4}
\end{align}

Substituting~\eqref{eq:app-info-4}, \eqref{eq:app-info-3}, and
\eqref{eq:app-info-2} into~\eqref{eq:app-info-1}, we can write
\begin{equation*}
  \pi_{t+1}(s_{t+1}) = F_{t+1}(\pi_t, \tilde \gamma^1_t, \tilde \gamma^2_t,
  z_{t+1})(s_{t+1})
\end{equation*}
where $F_{t+1}(\cdot)$ is given by~\eqref{eq:app-info-1}, \eqref{eq:app-info-2},
\eqref{eq:app-info-3}, and~\eqref{eq:app-info-4}.

\section{Proof of Proposition~\ref{prop:MDP}} \label{app:MDP}

Fix a coordinator strategy $\boldsymbol\psi$. Consider a realization
$\delta_{t+1}$ of the common information $\Delta_{t+1}$. Let $\pi_{1:t}$
be the corresponding realization of $\Pi_{1:t}$ and $(\tilde
\gamma^1_{1:t}, \tilde \gamma^2_{1:t})$ the corresponding choice of partial
functions until time $t$. Then, for any Borel subset $A \subset \PSP{\ALPHABET S}$, where
$\PSP{\ALPHABET S}$ is the space of probability mass functions over the finite set $\ALPHABET S$ (the space of realization of $S_t$),  we can write using
Proposition~\ref{prop:info}
\begin{align}
  \PR{\Pi_{t+1} \in A | \delta_t,\pi_{1:t}, \tilde \gamma^1_{1:t}, \tilde \gamma^2_{1:t}}
  = &\sum_{z_{t+1}} \IND_{A}(F_{t+1}(\pi_t, \tilde \gamma^1_t, \tilde
  \gamma^2_t,z_{t+1})) 
  \cdot \PR{Z_{t+1} = z_{t+1} | \delta_t, \pi_{1:t},\tilde \gamma^1_{1:t},\tilde
  \gamma^2_{1:t}} \label{eq:app-MDP-1}
\end{align}

Now, using~\eqref{eq:coordinator-observation}, we have
\begin{align}
  \PR{Z_{t+1} = z_{t+1} | \delta_t, \pi_{1:t}, \tilde \gamma^1_{1:t}, \tilde
  \gamma^2_{1:t}} 
  &= \sum_{s_t} \IND_{\hat h_t(s_t)}(z_{t+1})
  \cdot
  \PR{S_t = s_t | \delta_t,\pi_{1:t}, \tilde \gamma^1_{1:t}, \tilde
  \gamma^2_{1:t}} \notag \\
  &=  \sum_{s_t} \IND_{\hat h_t(s_t)}(z_{t+1})\cdot \pi_t(s_t) 
  \label{eq:app-MDP-2}
\end{align}

Substituting~\eqref{eq:app-MDP-2} back in~\eqref{eq:app-MDP-1}, we get
\begin{align}
  \PR{\Pi_{t+1} \in A | \delta_t,\pi_{1:t}, \tilde \gamma^1_{1:t}, \tilde \gamma^2_{1:t}}
  &= \sum_{z_{t+1}}\sum_{s_t} 
  \IND_{A}(F_{t+1}(\pi_t, \tilde \gamma^1_t, \tilde
  \gamma^2_t,z_{t+1}))  
   \cdot
   \IND_{\hat h_t(s_t)}(z_{t+1})\cdot \pi_t(s_t) \notag \\
  &= 
  \PR{\Pi_{t+1} \in A | \pi_{t}, \tilde \gamma^1_{t}, \tilde \gamma^2_{t}},
\end{align}
thereby proving~\eqref{eq:Markov_State}.

Now, using Proposition~\ref{prop:state} we can write, 
\begin{align}
  \hskip 2em & \hskip -2em 
    \EXP{\hat{c}_t(S_t,\gamma^1_t,\gamma^2_t,S_{t+1}) | 
      \delta_t,\pi_{1:t},\tilde{\gamma}^{1}_{1:t},\tilde{\gamma}^{2}_{1:t}} \notag \\
    &= 
  \smashoperator[l]{\sum_{s_t, v_t, w^1_{t+1}, w^2_{t+1}} }
  \hat c_t(s_t, \tilde{\gamma}^1_t, \tilde{\gamma}^2_t, 
          \hat f_{t+1}(s_t,v_t,w^1_{t+1},w^2_{t+1},
          \tilde\gamma^1_t, \tilde \gamma^2_t))
  \cdot \PR{V_t = v_t} 
   \notag \\
   &\quad \cdot
   \PR{W^1_{t+1} = w^1_{t+1}} \cdot \PR{W^2_{t+1} = w^2_{t+1}}
  \displaybreak[2]
  \cdot
   \PR{S_t = s_t |\delta_t, \pi_{1:t}, \tilde \gamma^1_{1:t}, \tilde \gamma^2_{1:t}} 
  \notag \\
  &=
  \smashoperator[l]{\sum_{s_t, v_t, w^1_{t+1}, w^2_{t+1}} }
  \hat c_t(s_t, \tilde\gamma^1_t, \tilde\gamma^2_t, 
          \hat f_{t+1}(s_t,v_t,w^1_{t+1},w^2_{t+1},
          \tilde\gamma^1_t, \tilde \gamma^2_t))
  \cdot
   \PR{V_t = v_t}  \notag \\
  &\quad \cdot \PR{W^1_{t+1} = w^1_{t+1}} \cdot \PR{W^2_{t+1} = w^2_{t+1}}
    \cdot \pi_t(s_t) \notag \\
  & \BYDEFINITION C_t(\pi_t, \tilde \gamma^1_t, \tilde \gamma^2_t).
\end{align}
This proves~\eqref{eq:MDP_Cost}.

\section{Piecewise linearity and concavity of value function} \label{app:convex}

Since $C_t(\Pi_t, \gamma^1_1,\gamma^2_t) =
\EXP{\hat{c}_t(S_t,\gamma^1_t,\gamma^2_t,S_{t+1})|\Delta_t,\Pi_{1:t},\gamma^{1}_{1:t},\gamma^{2}_{1:t}}$,
the value function  at time $T$ can be written as,
\begin{equation} \label{eq:appC.1}
  J_{T}(\pi) = \inf _{\tilde\gamma^1,\tilde\gamma^2} \EXP{{\hat c_T}(S_T,
  \tilde\gamma^1,\tilde\gamma^2,S_{T+1})|
  \Pi_T=\pi,\gamma^1_T=\tilde\gamma^1,\gamma^2_T=\tilde\gamma^2}.
\end{equation}
For a given choice of $\tilde\gamma^1,\tilde\gamma^2$, the expectation in
equation \eqref{eq:appC.1} can be written as:
\begin{align}
  \smashoperator[l]{\sum_{s_T, v_T, w^1_{T+1}, w^2_{T+1}} }
  &\hat c_T(s_T, \tilde\gamma^1, \tilde\gamma^2, 
          \hat f_{T+1}(s_T,v_T,w^1_{T+1},w^2_{T+1},
          \tilde\gamma^1, \tilde \gamma^2))\notag \\& \cdot\PR{V_T = v_T,W^1_{T+1} = w^1_{T+1},W^2_{T+1} = w^2_{T+1}} \cdot \pi(s_T) \label{eq:appC.2}
\end{align}
The expression in \eqref{eq:appC.2} is linear in $\pi$. Therefore, the value
function $J_T(\pi)$ is the infimum of finitely many linear functions of $\pi$.
Hence, $J_T(\pi)$ is a piecewise-linear (and hence concave) function. We now
proceed inductively.
  
First assume that $J_{t+1}(\pi)$ is a concave function. Then, $J_{t+1}$ can be
written as infimum of a family of affine functions.
\begin{equation}
  J_{t+1}(\pi) = \inf_i{\sum_{s \in \mathcal{S}}a_i(s)\cdot \pi(s)+b_i}, \label{eq:concave_1}
\end{equation}
where $a_i(s), s \in \mathcal{S}$ and $b_i$ are real numbers. The value function
at time $t$ is given as:
\begin{align}
  J_{t}(\pi) = \inf_{\tilde\gamma^1,\tilde\gamma^2} \big[&\EXP{{\hat c_t}(S_t,
    \tilde\gamma^1,\tilde\gamma^2,S_{t+1}) |
    \Pi_t=\pi,\gamma^1_t=\tilde\gamma^1,\gamma^2_t=\tilde\gamma^2} \notag\\
   &+ \EXP{J_{t+1}(\Pi_{t+1}) |
   \Pi_t=\pi ,\gamma^1_t=\tilde\gamma^1,\gamma^2_t=\tilde\gamma^2}\big]
   \label{eq:appC.4}
\end{align} 
For a given choice of $\tilde\gamma^1,\tilde\gamma^2$, the first expectation in
\eqref{eq:appC.4} can be written as
\begin{align}
  \smashoperator[l]{\sum_{s_t, v_t, w^1_{t+1}, w^2_{t+1}} }
  &\hat c_t(s_t, \tilde\gamma^1, \tilde\gamma^2, 
          \hat f_{t+1}(s_t,v_t,w^1_{t+1},w^2_{t+1},
          \tilde\gamma^1, \tilde \gamma^2)) \nonumber \\
   &\quad \cdot \PR{V_t = v_t,W^1_{t+1} = w^1_{t+1},W^2_{t+1} = w^2_{t+1}}
          \cdot \pi(s_t)
\end{align}        
Thus, for a given choice of $\tilde\gamma^1,\tilde\gamma^2$, the first
expectation in~\eqref{eq:appC.4} is linear in $\pi$. Using
Proposition~\ref{prop:info}, the  second expectation in~\eqref{eq:appC.4} can be
written as:
\begin{align}
   &\EXP{J_{t+1}(F_{t+1}(\Pi_t,\tilde\gamma^1,\tilde\gamma^2,Z_{t+1})) |
         \Pi_t=\pi,\gamma^1_t=\tilde\gamma^1,\gamma^2_t=\tilde\gamma^2} 
         \notag\\
   &=
    \sum_{z_{t+1}} J_{t+1}(F_{t+1}(\pi,\tilde\gamma^1,\tilde\gamma^2,z_{t+1})) 
     \cdot
     \PR{Z_{t+1} =z_{t+1}|\Pi_t =\pi,\gamma^1_t=\tilde\gamma^1,
          \gamma^2_t=\tilde\gamma^2} \notag \\
   &= 
   \sum_{z_{t+1}}\bigg[
   \inf_i\Big\{\sum_{s}a_i(s) \cdot
    (F_{t+1}(\pi,\tilde\gamma^1,\tilde\gamma^2,z_{t+1}))(s) 
   + b_i\Big\} \bigg]  \cdot
   \PR{Z_{t+1} =z_{t+1}|\Pi_t =\pi,\gamma^1_t=\tilde\gamma^1,\gamma^2_t=\tilde\gamma^2} 
   \label{eq:appC.5}   
\end{align}
We now focus on each term in the outer summation in \eqref{eq:appC.5}. For each
value of $z_{t+1}$, these terms can be written as:
\begin{align}
   \inf_i\Big\{&
   \sum_{s}a_i(s)\cdot (F_{t+1}(\pi,\tilde\gamma^1,\tilde\gamma^2,z_{t+1}))(s)   
   \cdot
   \PR{Z_{t+1} =z_{t+1}|\Pi_t =\pi,\gamma^1_t=\tilde\gamma^1,
      \gamma^2_t=\tilde\gamma^2} \notag \\ 
    &+ b_i \cdot \PR{Z_{t+1} =z_{t+1}|\Pi_t =\pi,\gamma^1_t=\tilde\gamma^1,
    \gamma^2_t=\tilde\gamma^2}\Big\}
    \label{eq:concavity_pf_1}
\end{align}
We first note that the term $b_i \cdot \PR{Z_{t+1} =z_{t+1}|\Pi_t
=\pi,\gamma^1_t=\tilde\gamma^1,\gamma^2_t=\tilde\gamma^2}$ is affine in $\pi$.
This is because:
\begin{align}
    b_i \cdot \PR{Z_{t+1} =z_{t+1}|\Pi_t =\pi,\gamma^1_t=\tilde\gamma^1,\gamma^2_t=\tilde\gamma^2} 
    &= b_i \cdot \sum_{s' \in \mathcal{S}} \IND_{\hat{h}_t(s')}(z_{t+1})\cdot \pi(s') \label{eq:concavity_pf_1.1}
   \end{align}
Moreover, using the characterization of $F_{t+1}$ from the proof of Proposition
\ref{prop:info} (Appendix~\ref{app:info}), we can write the term with
coefficients $a_i(s)$ in~\eqref{eq:concavity_pf_1} as 
\begin{align}
   a_i(s) \cdot \bigg\{&\sum_{s_t, v_t, w^1_{t+1}, w^2_{t+1}} 
  \IND_{s}( \hat f_{t+1}
    (s_t, v_t, w^1_{t+1}, w^2_{t+1}, \tilde \gamma^1_t, \tilde \gamma^2_t))
  \notag \\
  & \cdot 
   \PR{V_t = v_t,W^1_{t+1} = w^1_{t+1},W^2_{t+1} = w^2_{t+1}} 
    \cdot \IND_{\hat{h}(s_t)}(z_{t+1}) \pi(s_t)\bigg\}
    \label{eq:concavity_pf_2}
\end{align}
which is also affine in $\pi$. Using equations
\eqref{eq:concavity_pf_1}, \eqref{eq:concavity_pf_1.1} and~\eqref{eq:concavity_pf_2} in \eqref{eq:appC.5}, we conclude that for a given
choice of $\tilde\gamma^1, \tilde\gamma^2$, the second expectation in
\eqref{eq:appC.4} is concave in $\pi$. Thus, the value function $J_t(\pi)$ is
the minimum of finitely many functions each of which is the sum of an affine and
a concave function of $\pi$. This implies that $J_t$ is concave in $\pi$. This
completes the induction argument.

\section{Proof of Proposition~\ref{prop:equiv_info}} \label{proof:equiv_info}

\begin{enumerate}
  \item Recall that $Z_{t+1}= (Y^1_{t-n+1},Y^2_{t-n+1},U^1_{t-n+1},U^2_{t-n+1})$
    and $\Delta_{t+1} = \Delta_t \cup Z_{t+1}$. Fix a coordination strategy
    $\boldsymbol \psi$ and consider a realization $\delta_{t+1}$ of
    $\Delta_{t+1}$. Then,  
    \begin{align}
      \theta_{t+1}(x_{t-n+1}) &\DEFINED \PPR(X_{t-n+1}=x_{t-n+1}|
        \delta_{t+1}) \nonumber \\
      &= \PPR(X_{t-n+1}=x_{t-n+1}|
        \delta_{t},y^1_{t-n+1},y^2_{t-n+1},u^1_{t-n+1},u^2_{t-n+1}) \nonumber\\
      &= \sum_{x \in \mathcal{X}} 
      \PPR(X_{t-n+1}=x_{t-n+1}|X_{t-n}=x,u^1_{t-n+1},u^2_{t-n+1})  \\
      &\quad\cdot 
      \PPR(X_{t-n}=x|\delta_{t},y^1_{t-n+1},y^2_{t-n+1},u^1_{t-n+1},
            u^2_{t-n+1}) \nonumber\\
      &= \sum_{x \in \mathcal{X}} 
      \PPR(X_{t-n+1}=x_{t-n+1}|X_{t-n}=x,u^1_{t-n+1},u^2_{t-n+1})\\ 
      &\quad\cdot  
      \frac{\PPR(X_{t-n}=x,y^1_{t-n+1},y^2_{t-n+1},u^1_{t-n+1},u^2_{t-n+1}|\delta_t)}
      {\sum_{x'}\PPR(X_{t-n}=x',y^1_{t-n+1},y^2_{t-n+1},u^1_{t-n+1},u^2_{t-n+1}|\delta_t)} 
      \label{eq:appE.1}
    \end{align} 
    Consider the second term of~\eqref{eq:appE.1}, and note that under any
    coordination strategy $\boldsymbol\psi$, the variables
    $u^1_{t-n+1},u^2_{t-n+1}$ are deterministic functions of
    $y^1_{t-n+1},y^2_{t-n+1}$ and $\delta_t$ (which is same as $y^{1:2}_{1:t-n},
    u^{1:2}_{1:t-n}$). Therefore,  the second term of~\eqref{eq:appE.1} can be
    written as 
    \begin{multline}
      \frac{\PPR^{\boldsymbol\psi}(u^1_{t-n+1},u^2_{t-n+1}|
        y^1_{t-n+1},y^2_{t-n+1},\delta_t)
        \cdot
        \PPR(y^1_{t-n+1},y^2_{t-n+1}|X_{t-n}=x)
        \cdot 
        \PPR(X_{t-n}=x|\delta_t)}
      {\sum_{x'}\PPR^{\boldsymbol\psi}(u^1_{t-n+1},u^2_{t-n+1}|
             y^1_{t-n+1},y^2_{t-n+1},\delta_t)
            \cdot 
            \PPR(y^1_{t-n+1},y^2_{t-n+1}|X_{t-n}=x')
            \cdot
            \PPR(X_{t-n}=x'|\delta_t)} 
      \\
      = \frac{\PPR(y^1_{t-n+1}|X_{t-n}=x)
          \cdot
          \PPR(y^2_{t-n+1}|X_{t-n}=x)
          \cdot
          \theta_t(x)}
      {\sum_{x'}\PPR(y^1_{t-n+1}|X_{t-n}=x)
          \cdot
          \PPR(y^2_{t-n+1}|X_{t-n}=x)
          \cdot
          \theta_t(x')}
      \label{eq:appE.2}
    \end{multline}
    Substituting~\eqref{eq:appE.2} in~\eqref{eq:appE.1}, we conclude that
    $\theta_{t+1}$ is a function of $\theta_t$ and  $z_{t+1}$.

    Consider next $r^k_{t+1}\DEFINED (r^k_{m,(t+1)}, t-n+2\leq m \leq t)$. For
    $m=t$, we have $r^k_{t,(t+1)}\DEFINED\gamma^k_t(\cdot,Y^k_{t-n+1})$. Since
    $Y^k_{t-n+1}$ is a part of $Z_{t+1}$, therefore $r^k_{t,(t+1)}$ is a
    function of $\gamma^k_t$ and $Z_{t+1}$. Also, for $m = t-n+2,
    t-n+3,\ldots,t-1$,
    \begin{align}
      r^k_{m,t+1}(\cdot) &\DEFINED 
      \gamma^k_m(\cdot, Y^{k}_{m-n+1:t+1-n}, U^{k}_{m-n+1:t+1-n}) \nonumber \\
      &= 
      \gamma^k_m(\cdot,Y^k_{t-n+1},U^k_{t-n+1}, Y^{k}_{m-n+1:t-n}, U^{k}_{m-n+1:t-n}) 
      \nonumber \\
      &=  r^k_{m,t}(\cdot,Y^k_{t-n+1},U^k_{t-n+1})
    \end{align}
    Thus, for $m = t-n+2, t-n+3,\ldots,t-1$, $r^k_{m,t+1}$ is a function of
    $r^k_{m,t}$ and $Z_{t+1}$.

  \item We will first show that the coordinator's belief $\Pi_t$ defined
    in~\eqref{eq:define_pi} is a function of $(\Theta_t,r^1_t,r^2_t)$. That is,
    there exist functions $H_t$, for $t=1,2,\ldots,T$, such that
    \begin{equation} \label{eq:H_function}
      \Pi_t = H_t(\Theta_t,r^1_t,r^2_t) 
    \end{equation}
    Using this fact with using equation~\eqref{eq:MDP_Cost} from
    Proposition~\ref{prop:MDP}, we can conclude that
    \begin{align}
      \EXP{\hat{c}_t(S_t,\gamma^1_t,\gamma^2_t,S_{t+1})|
      \Delta_t,\Pi_{1:t},\gamma^{1}_{1:t},\gamma^{2}_{1:t}}
      &=  C_t(\Pi_t, \gamma^1_1,\gamma^2_t) \nonumber \\
      &= \hat{C}_t(\Theta_t,r^1_t,r^2_t,\gamma^1_1,\gamma^2_t) 
      \label{eq:appEpf1}
    \end{align}  
    where we use the fact that $\Pi_t$ is a function of $(\Theta_t,r^1_t,r^2_t)$
    in equation~\eqref{eq:appEpf1}. In order to prove~\eqref{eq:H_function}, we
    need the following lemma:  

    \begin{lemma}
      $S_t \DEFINED (X_{t-1},\Lambda^{1}_{t},\Lambda^{2}_{t})$ is a
      deterministic function of $(X_{t-n},V_{t-n+1:t-1}$,$W^1_{t-n+1:t}$,
      $W^2_{t-n+1:t},r^1_t,r^2_t)$. That is, there exists a fixed deterministic
      function $D_t$ such that
      \begin{equation}
        S_t \DEFINED (X_{t-1},\Lambda^{1}_{t},\Lambda^{2}_{t}) = 
        D_t(X_{t-n},V_{t-n+1:t-1},W^1_{t-n+1:t},W^2_{t-n+1:t},r^1_t,r^2_t) 
      \end{equation}
    \end{lemma}

    \begin{proof} 
      We can reconstruct $(X_{t-n+1:t-1},\Lambda^{1}_{t},\Lambda^{2}_{t})$ from
      $(X_{t-n},V_{t-n+1:t-1},W^1_{t-n+1:t}$,\\ $W^2_{t-n+1:t},r^1_t,r^2_t)$
      using the given dynamics of the system~\eqref{eq:dynamics}, the
      observation equation~\eqref{eq:obs} and the definition of $r^k_t$ in a
      straight forward manner. Firstly note that
      \begin{equation}
        (X_{t-n+1:t-1},\Lambda^{1}_{t},\Lambda^{2}_{t}) 
        =  (X_{t-n+1:t-1},Y^{1:2}_{t-n+1:t},U^{1:2}_{t-n+1:t-1})
      \end{equation}
      We first look at the random variables
      $(X_{t-n+1},Y^{1:2}_{t-n+1},U^{1:2}_{t-n+1})$. We have, for $k=1,2$,
      \begin{align}
        Y^k_{t-n+1} &= h^k_{t-n+1}(X_{t-n},W^k_{t-n+1}) \nonumber \\
        U^k_{t-n+1} &= r^k_{t-n+1,t}(Y^k_{t-n+1}) \nonumber \\
      \end{align}
      Further, by the system dynamics,
      \begin{align}
        X_{t-n+1} = f_t(X_{t-n},U^{1:2}_{t-n+1},V_{t-n+1})
      \end{align}
      Thus $(X_{t-n+1},Y^{1:2}_{t-n+1},U^{1:2}_{t-n+1})$ is a deterministic
      function of $(X_{t-n},W^{1:2}_{t-n+1},V_{t-n+1},r^{1:2}_{t-n+1,t})$. Now
      assume $(X_{t-n+1:m},Y^{1:2}_{t-n+1:m},U^{1:2}_{t-n+1:m})$ is a function
      of $(X_{t-n},W^{1:2}_{t-n+1:m},V_{t-n+1:m},$ $r^{1:2}_{t-n+1:m,t})$. We
      have shown above that this is true for $m=t-n+1$. Then, for $m=t-n+1:t-2$,
      \begin{align}
         Y^k_{m+1} &= h^k_{m+1}(X_{m},W^k_{m+1}) \nonumber \\
         U^k_{m+1} &= r^k_{m+1,t}(Y^k_{t-n+1:m+1},U^k_{t-n+1:m}) \nonumber
      \end{align}
      Further, by the system dynamics,
      \begin{align}
        X_{m+1} = f_t(X_{m},U^{1:2}_{m+1},V_{m+1})
      \end{align}
      Thus, $(X_{m+1},Y^{1:2}_{m+1},U^{1:2}_{m+1})$ is a deterministic function
      of
      \begin{equation*}
        (X_m,Y^{1:2}_{t-n+1:m},U^{1:2}_{t-n+1:m},
            W^{1:2}_{m+1},V_{m+1},r^{1:2}_{m+1,t})
      \end{equation*}
      Combining this with our induction hypothesis, we conclude that
      $(X_{t-n+1:m+1},Y^{1:2}_{t-n+1:m+1}$,\\$U^{1:2}_{t-n+1:m+1})$ is a
      function of
      $(X_{t-n},W^{1:2}_{t-n+1:m+1},V_{t-n+1:m+1},r^{1:2}_{t-n+1:m+1,t})$. Thus,
      by induction we have that
      \[(X_{t-n+1:t-1},Y^{1:2}_{t-n+1:t-1},U^{1:2}_{t-n+1:t-1})\]
      is a function of 
      \[(X_{t-n},W^{1:2}_{t-n+1:t-1},V_{t-n+1:t-1},r^{1:2}_{t-n+1:t-1,t})\]
      Finally noting that $Y^k_t = h^k_t(X_{t-1},W^k_t)$ and that $r^k_t =
      r^k_{t-n+1:t-1,t}$, we can conclude that there exists a deterministic
      function $\hat{D}_t$ such that
      \begin{equation}
        (X_{t-n+1:t-1},Y^{1:2}_{t-n+1:t},U^{1:2}_{t-n+1:t-1}) = 
        \hat{D}_t(X_{t-n},V_{t-n+1:t-1},W^1_{t-n+1:t},W^2_{t-n+1:t},r^1_t,r^2_t) 
      \end{equation}
      This implies the existence of functions $D_t$ such that
      \begin{equation}
        S_t \DEFINED (X_{t-1},\Lambda^{1}_{t},\Lambda^{2}_{t}) = 
         D_t(X_{t-n},V_{t-n+1:t-1},W^1_{t-n+1:t},W^2_{t-n+1:t},r^1_t,r^2_t) 
      \end{equation}
  \end{proof}

  Now consider
  \begin{align}
    &\Pi_t(s_t)\DEFINED 
    \PR^{\boldsymbol\psi}{S_t = s_t | \Delta_t, \gamma^1_{1:t-1},\gamma^2_{1:t-1}} 
    \nonumber \\
    &\quad=
    \sum_{\substack{x_{t-n},v_{t-n+1:t-1},\\w^{1:2}_{t-n+1:t},\\\tilde{r}^1_t,\tilde{r}^2_t}}
    \IND_{s_t}\{D_t(x_{t-n},v_{t-n+1:t-1},w^{1:2}_{t-n+1:t},\tilde{r}^1_t,\tilde{r}^2_t)\}
    \notag \\
    &\qquad\cdot 
    \PR{x_{t-n},v_{t-n+1:t-1},w^{1:2}_{t-n+1:t},\tilde{r}^1_t,\tilde{r}^2_t|
        \Delta_t, \gamma^1_{1:t-1},\gamma^2_{1:t-1}}
    \label{eq:appE.2.1}
  \end{align}
  Note that $r^1_t,r^2_t$ are completely determined by $\Delta_t$ and
  $\gamma^{1:2}_{1:t-1}$ and the noise random variables
  $v_{t-n+1:t-1},w^{1:2}_{t-n+1:t}$ are independent of the conditioning terms
  and $X_{t-n}$. We can therefore write~\eqref{eq:appE.2.1} as
  \begin{align}
    &\sum \IND_{s_t}\{D_t(x_{t-n},v_{t-n+1:t-1},w^{1:2}_{t-n+1:t},\tilde{r}^1_t,\tilde{r}^2_t)\} \cdot \PR{v_{t-n+1:t-1},w^{1:2}_{t-n+1:t}} \notag \\
    &\quad \cdot 
    \IND_{\tilde{r}^1_t,\tilde{r}^2_t}(r^1_{t},r^2_{t}) 
	 \cdot
    \PR{x_{t-n}|\Delta_t, \gamma^1_{1:t-1},\gamma^2_{1:t-1}}
    \label{eq:appE.2.2}
  \end{align} 
  
  In the last term of~\eqref{eq:appE.2.2}, we can drop $\gamma^{1:2}_{1:t-1}$
  from the conditioning terms since they are functions of $\Delta_t$. The last
  term is therefore same as $\PR{x_{t-n}|\Delta_t} = \Theta_t$. Thus, $\Pi_t$ is
  a function of $\Theta_t$ and $r^1_t,r^2_t$. 
  
  \item  Consider the following probability:
    \begin{align}
      &\PR{\Theta_{t+1}=\theta_{t+1},r^1_{t+1}=\tilde{r}^1_{t+1},
           r^2_{t+1}=\tilde{r}^2_{t+1}|\delta_t,\theta_{1:t},
           \tilde{\gamma}^{1:2}_{1:t},\tilde{r}^1_{1:t},\tilde{r}^2_{1:t}} 
       \nonumber \\
     &\quad= \sum_{z_{t+1}} 
     \IND_{\theta_{t+1}}(Q_{t+1}(\theta_t,z_{t+1}))
     \cdot 
     \IND_{\tilde{r}^1_{t+1}}(Q^1_{t+1}(\tilde{r}^1_t, 
           \tilde \gamma^1_t,z_{t+1})) \nonumber \\
     &\qquad\cdot
     \IND_{\tilde{r}^2_{t+1}}(Q^2_{t+1}(\tilde{r}^2_t, 
           \tilde \gamma^2_t,z_{t+1}))
    \cdot 
    \PR{Z_{t+1}=z_{t+1}|\delta_t,\tilde{\gamma}^{1:2}_{1:t},
        \tilde{r}^1_{1:t},\tilde{r}^2_{1:t}} 
     \label{eq:appE.3.1} 
   \end{align}
   The probability in equation~\eqref{eq:appE.3.1} can be written as:
   \begin{align}
     \hskip 1em & \hskip -1em 
     \PR{Z_{t+1} = z_{t+1} |
     \delta_t,\tilde{\gamma}^{1:2}_{1:t},\tilde{r}^1_{1:t},\tilde{r}^2_{1:t}}
     \notag \\
     \displaybreak[2]
     &= \sum_{s_t} \IND_{\hat h_t(s_t)}(z_{t+1})
     \cdot
     \PR{S_t = s_t |\delta_t,\tilde{\gamma}^{1:2}_{1:t},\tilde{r}^1_{1:t},\tilde{r}^2_{1:t}} \notag\\
          \displaybreak[2]
     &= \sum_{s_t} \IND_{\hat h_t(s_t)}(z_{t+1})
     \cdot
     \PR{S_t = s_t |\delta_t} \notag\\
     &= \sum_{s_t} \IND_{\hat h_t(s_t)}(z_{t+1})
     \cdot \pi_t(s_t) \notag\\
     &= \sum_{s_t} \IND_{\hat h_t(s_t)}(z_{t+1}) 
     \cdot 
     H_t(\theta_t,\tilde{r}^1_t,\tilde{r}^2_t)(s_t)  \label{eq:appE.3.2}
   \end{align}
   Substituting~\eqref{eq:appE.3.2} back in~\eqref{eq:appE.3.1}, we get
   \begin{align}
     &\PR{\Theta_{t+1}=\theta_{t+1}, r^1_{t+1}=\tilde{r}^1_{t+1},
     r^2_{t+1}=\tilde{r}^2_{t+1} |
     \delta_t,\theta_{1:t}, \tilde{\gamma}^{1:2}_{1:t},
     \tilde{r}^1_{1:t},\tilde{r}^2_{1:t}} \nonumber \\
     &\quad=\sum_{z_{t+1},s_t} 
     \IND_{\theta_{t+1}}(Q_{t+1}(\theta_t,z_{t+1}))
     \cdot
     \IND_{\tilde{r}^1_{t+1}}(Q^1_{t+1}(\tilde{r}^1_t,
           \tilde \gamma^1_t,z_{t+1})) \nonumber \\
     &\qquad\cdot 
     \IND_{\tilde{r}^2_{t+1}}(Q^2_{t+1}(\tilde{r}^2_t, 
           \tilde \gamma^2_t,z_{t+1})) 
     \cdot
     \IND_{\hat h_t(s_t)}(z_{t+1})
     \cdot
     H_t(\theta_t,\tilde{r}^1_t,\tilde{r}^2_t)(s_t) \notag\\
     &\quad= 
     \PR{\Theta_{t+1}=\theta_{t+1},r^1_{t+1}=\tilde{r}^1_{t+1},
          r^2_{t+1}=\tilde{r}^2_{t+1}|\theta_{t},\tilde{r}^1_{t},
          \tilde{r}^2_{t},\tilde{\gamma}^1_{t},\tilde{\gamma}^2_{t}}
  \end{align}
  thereby proving~\eqref{eq:Markov_State_2}.
\end{enumerate}

\section*{Acknowledgments}
This research was supported in part by NSF Grant CCR-0325571 and NASA Grant
NNX06AD47G.



\begin{thebibliography}{10}
\providecommand{\url}[1]{#1}
\csname url@samestyle\endcsname
\providecommand{\newblock}{\relax}
\providecommand{\bibinfo}[2]{#2}
\providecommand{\BIBentrySTDinterwordspacing}{\spaceskip=0pt\relax}
\providecommand{\BIBentryALTinterwordstretchfactor}{4}
\providecommand{\BIBentryALTinterwordspacing}{\spaceskip=\fontdimen2\font plus
\BIBentryALTinterwordstretchfactor\fontdimen3\font minus
  \fontdimen4\font\relax}
\providecommand{\BIBforeignlanguage}[2]{{%
\expandafter\ifx\csname l@#1\endcsname\relax
\typeout{** WARNING: IEEEtran.bst: No hyphenation pattern has been}%
\typeout{** loaded for the language `#1'. Using the pattern for}%
\typeout{** the default language instead.}%
\else
\language=\csname l@#1\endcsname
\fi
#2}}
\providecommand{\BIBdecl}{\relax}
\BIBdecl

\bibitem{KumarVaraiya:1986}
P.~R. Kumar and P.~Varaiya, \emph{Stochastic Systems: Estimation Identification
  and Adaptive Control}.\hskip 1em plus 0.5em minus 0.4em\relax Prentice Hall,
  1986.

\bibitem{Witsenhausen:1971}
H.~S. Witsenhausen, ``Separation of estimation and control for discrete time
  systems,'' \emph{Proc. {IEEE}}, vol.~59, no.~11, pp. 1557--1566, Nov. 1971.

\bibitem{WalrandVaraiya:1978}
P.~Varaiya and J.~Walrand, ``On delayed sharing patterns,'' \emph{{IEEE} Trans.
  Autom. Control}, vol.~23, no.~3, pp. 443--445, 1978.

\bibitem{Kurtaran:1979}
B.~Kurtaran, ``Corrections and extensions to "decentralized stochastic control
  with delayed sharing information pattern",'' \emph{{IEEE} Trans. Autom.
  Control}, vol.~24, no.~4, pp. 656--657, Aug. 1979.

\bibitem{Witsenhausen:1976}
H.~S. Witsenhausen, ``Some remarks on the concept of state,'' in
  \emph{Directions in Large-Scale Systems}, Y.~C. Ho and S.~K. Mitter,
  Eds.\hskip 1em plus 0.5em minus 0.4em\relax Plenum, 1976, pp. 69--75.

\bibitem{Zhang:2009}
H.~Zhang, ``Partially observable markov decision processes: A geometric
  technique and analysis,'' \emph{Operations Research}, 2009.

\bibitem{Aicardi:1987}
M.~Aicardi, F.~Davoli, and R.~Minciardi, ``Decentralized optimal control of
  markov chains with a common past information set,'' \emph{IEEE Transactions
  on Automatic Control}, vol.~32, no.~11, Nov. 1987.

\bibitem{Kurtaran:1976}
B.~Kurtaran, ``Decentralized stochastic control with delayed sharing
  information pattern,'' \emph{{IEEE} Trans. Autom. Control}, vol.~21, pp.
  576--581, Aug. 1976.

\bibitem{NayyarTeneketzis:2008}
A.~Nayyar and D.~Teneketzis, ``On the structure of real-time encoders and
  decoders in a multi-terminal communication system,'' \emph{IEEE Trans. Info.
  Theory}, 2009, submitted.

\bibitem{MahajanNayyarTeneketzis:2008}
A.~Mahajan, A.~Nayyar, and D.~Teneketzis, ``Identifying tractable decentralized
  control problems on the basis of information structures,'' in
  \emph{proceedings of the 46th {A}llerton conference on communication, control
  and computation}, Sep. 2008, pp. 1440--1449.

\end{thebibliography}
\end{document}